\documentclass[journal]{IEEEtran}
\IEEEoverridecommandlockouts
\usepackage{epsfig}
\usepackage{epstopdf} 
\usepackage{mathrsfs}
\usepackage[noadjust]{cite}
\usepackage{graphicx,color,overpic,psfrag}
\usepackage{amsmath, amssymb}
\usepackage{latexsym}
\usepackage{bm}
\usepackage{amssymb}
\usepackage{cases}
\usepackage{array}
\usepackage{fancyhdr}
\usepackage{setspace}
\usepackage{subfigure}
\usepackage{caption}
\UseRawInputEncoding
\usepackage{indentfirst}
\usepackage{cases}
\usepackage{url}
\usepackage{algpseudocode}
\usepackage{algorithm}
\usepackage{blkarray}
\usepackage{booktabs}
\usepackage{multirow}
\usepackage{dsfont}
\usepackage{tabularx}
\usepackage[table]{xcolor}
\usepackage{amsfonts}
\usepackage{amsthm}
\usepackage{stfloats}
\usepackage{letltxmacro}
\usepackage{cuted}
\usepackage{cellspace}

\newcommand{\ra}[1]{\renewcommand{\arraystretch}{#1}}

\newcommand{\figref}[1]{Fig. \ref{#1}}
\newcommand{\tabref}[1]{Table \ref{#1}}

\graphicspath{{figure/}}
\allowdisplaybreaks
\theoremstyle{remark}
\newtheorem{theorem}{Proposition}

\begin{document}
	
	\title{{M}assive {MIMO}-{OFDM} {C}hannel {A}cquisition with {M}ulti-group {A}djustable {P}hase {S}hift {P}ilots}
	
	\author{ Yu~Zhao,~\IEEEmembership{Graduate Student~Member,~IEEE,}
		Li~You,~\IEEEmembership{Senior~Member,~IEEE,}\\
		Jinke~Tang,~\IEEEmembership{Student~Member,~IEEE,}
		Mengyu~Qian,~\IEEEmembership{Graduate Student~Member,~IEEE,}\\
		Bin~Jiang,~\IEEEmembership{Member,~IEEE,}
		Xiang-Gen~Xia,~\IEEEmembership{Fellow,~IEEE,}
		and Xiqi~Gao,~\IEEEmembership{Fellow,~IEEE}
		
		\thanks{
			Part of this work has been accepted for presentation at the 2025 IEEE 101th Vehicular Technology Conference \cite{YZLY2025}.
				
			Yu Zhao, Li You, Jinke Tang, Mengyu Qian, Bin Jiang, and Xiqi Gao are with the National Mobile Communications Research Laboratory, Southeast University, Nanjing 210096, China, and also with the Purple Mountain Laboratories, Nanjing 211100, China (e-mail: yu\_zhao@seu.edu.cn, lyou@seu.edu.cn, jktang@seu.edu.cn, qianmy@seu.edu.cn, bjiang@seu.edu.cn, xqgao@seu.edu.cn). \textit{(Corresponding author: Li You.)}
			
			Xiang-Gen Xia is with the Department of Electrical and Computer Engineering, University of Delaware, Newark, DE 19716, USA (e-mail: xianggen@udel.edu). 
			}
	}

	\maketitle
	
	\begin{abstract}
		
		Massive multiple-input multiple-output - orthogonal frequency division multiplexing (MIMO-OFDM) systems face the challenge of high channel acquisition overhead while providing significant spectral efficiency (SE). Adjustable phase shift pilots (APSPs) are an effective technique to acquire channels with low overhead by exploiting channel sparsity. In this paper, we extend it to multiple groups and propose multi-group adjustable phase shift pilots (MAPSPs) to improve SE further. We first introduce a massive MIMO-OFDM system model and transform the conventional channel model in the space-frequency domain to the angle-delay domain, obtaining a sparse channel matrix. Then, we propose a method of generating MAPSPs through multiple basic sequences and investigate channel estimation processes. By analyzing the components of pilot interference, we elucidate the underlying mechanism by which interference affects MMSE estimation. Building upon this foundation, we demonstrate the benefit of phase scheduling in MAPSP channel estimation and establish the optimal design condition tailored for scheduling. Furthermore, we propose an implementation scheme based on Zadoff-Chu sequences that includes received signal pre-processing and pilot scheduling methods to mitigate pilot interference. Simulation results indicate that the MAPSP method achieves a lower mean square error (MSE) of estimation than APSP and significantly enhances SE in mobility scenarios.
		
	\end{abstract}
	
	
	\section{Introduction}
	As global mobile data traffic continues to surge, user capacity has become a central focus in the development of the next generation of wireless technology, specifically 6G \cite{HCLY2025,ericsson2024report,ZJLY2025}. In comparison to traditional multiple-input multiple-output (MIMO), massive MIMO technology further enhances the number of antennas. This increase in antennas directly correlates with the data transmission rate in the same band \cite{TLM2015,LYXC2020,YZhuLY2025}. In the future, 6G is expected to operate at higher frequencies and wider bandwidths while emphasizing the channel characteristics of space, delay, and frequency. These advantages enable 6G to achieve higher spectral efficiency (SE) and peak data rates, and make massive MIMO a fundamental technology for 6G \cite{WCXL2023}. 
	
	Orthogonal frequency division multiplexing (OFDM) is a modulation technique designed to effectively combat frequency-selective fading and intersymbol interference \cite{prasad2004ofdm}. Numerous studies have demonstrated that massive MIMO-OFDM technology can leverage the benefits of both MIMO and OFDM, resulting in enhanced throughput, SE, and reliability \cite{SHLL2024,JZJZ2022}. The implementation of spatial multiplexing within massive MIMO-OFDM significantly boosts system capacity \cite{TMRT2022}, but places a heavy reliance on precise channel state information (CSI) \cite{XNLY2024}. In high-mobility environments, the overhead of acquiring CSI is unacceptable, even when taking advantage of the reciprocity of uplink (UL) and downlink (DL) in time-division duplex (TDD) mode \cite{LYXQ2024}.
	
	Most channel acquisition methods rely on the pilot, and the phase shift orthogonal pilot (PSOP) is a commonly employed method for CSI acquisition in massive MIMO-OFDM systems \cite{dahlman20134g,XLWW2023}. While orthogonal pilot effectively prevents pilot interference, it incurs substantial pilot overhead, which is impractical in high-mobility scenarios requiring frequent pilot transmissions. As research on channel sparsity advances, many channel acquisition methods based on channel sparsity have emerged. Compressed sensing (CS) algorithm provides an effective strategy for obtaining CSI at low pilot overhead by channel sparsity \cite{TLNN2022,YZGS2022,YLAS2025}, but it is often limited by the computational complexity and the sparsity characteristics of the channel matrix. Compared with CS, pilot reuse provides a computationally simple and low overhead channel acquisition method \cite{LRMJ2024,LYXG2015,YCLY2020,LYMX2020}. Since transmitting the same pilot leads to interference at the receiving side, which negatively impacts the accuracy of CSI acquisition \cite{SSYC2020}, pilot scheduling plays a crucial role. By virtue of the concentrated properties of channel power in the finite delay and angle domain \cite{LYXG2015}, pilot interference can be mitigated or even eliminated through effective pilot scheduling \cite{XXBJ2016}. Building on the PSOP approach, adjustable phase shift pilots (APSP) are proposed as a more flexible non-orthogonal pilot reuse method \cite{YCLY2021,you2015channel}, which fully leverages channel sparsity through phase scheduling. In the space-frequency-time domain, also known as the triple-beam domain, deploying time-frequency phase shifted pilots over multiple OFDM symbols can also utilize sparsity to significantly improve SE \cite{JTXG2024}. Currently, most research aimed at reducing the pilot overhead in massive MIMO-OFDM systems focuses on leveraging the inherent sparsity of the channel, but neglects the coordination between pilot structure and channel characteristics. 
	
	With the rapid development of communication technology, relevant application domains and scenarios have become increasingly diversified and complex. To improve the performance of wireless systems in such intricate environments, more refined channel modeling is imperative. Recent studies have not only focused on the envelope characteristics of received signals but also emphasized the statistical properties of the phase. Contrary to the conventional assumption of a uniform distribution, the authors of \cite{JWSL2019} have highlighted that phase non-uniformity arises when the dominant component fluctuates. The channel model proposed in \cite{JWSL2023} effectively characterizes the non-uniformity of phase caused by propagation complexity. Moreover, field measurement data from emerging wireless application scenarios, such as body area networks, vehicle-to-vehicle communications, and unmanned aerial vehicle communications, have also exhibited multimodal characteristics in first-order statistics \cite{SL2014,DWJF2011,WKIG2016}. Consequently, developing an effective channel estimation method focusing on complex propagation environments in emerging applications has become crucial given the current technological landscape.
	\definecolor{lightblue}{rgb}{0.93,0.95,1.0}
	\newcolumntype{L}{>{\hspace*{-\tabcolsep}}l}
	\newcolumntype{R}{Sc<{\hspace*{-\tabcolsep}}}
	\begin{table}[h!]
		\captionsetup{font=footnotesize}
		\caption{List Of Notations}\label{tb:notation}
		\centering
		\ra{1.5}
		\scriptsize
		
		\begin{tabular}{LR}
			\toprule
			\textbf{Notation} &  \textbf{Definition} \\\rowcolor{lightblue}
			\midrule
			$\mathbf{G}_{k_q,\ell}$ &\parbox{7cm}{\centering The space-frequency domain channel response matrix of the $k_q\text{th}$ UT over the $\ell\text{th}$ OFDM symbol}\\ 
			\hline
			$\mathbf{H}_{k_q,\ell}$ & \parbox{7cm}{\centering The angle-delay domain channel response matrix of the $k_q\text{th}$ UT at the $\ell$ symbol} \\ \rowcolor{lightblue}
			\hline
			$\bar{\mathbf{H}}_{k_{q},\ell}$ & \parbox{7cm}{\centering The complement-0 extension of $\mathbf{H}_{k_{q},\ell}$}\\
			\hline
			$\mathbf{H}_{k_{q},\ell}^{\varphi_{k_{q}}-\varphi_{k'_{q}}}$ &\raisebox{0.4ex}{\parbox{7cm}{\centering The pilot interference of the $k'_{q}\text{th}$ UT from the $k_{q}\text{th}$ UT}}\\ \rowcolor{lightblue}
			\hline
			$\mathbf{A}$ &\parbox{7cm}{\centering The array response vector} \\
			\hline
			$\mathbf{P}_{k_q}$ &\parbox{7cm}{\centering The angle-delay domain channel power matrix of the $k_q\text{th}$ UT}\\ \rowcolor{lightblue}
			\hline
			$\bar{\mathbf{P}}_{k_{q}}$ &\parbox{7cm}{\centering The complement-0 extension of $\mathbf{P}_{k_{q}}$}\\
			\hline
			$\mathbf{P}_{k_{q}}^{\varphi_{k_{q}}-\varphi_{k'_{q}}}$ &\raisebox{0.5ex}{\parbox{7cm}{\centering The pilot interference power of the $k'_{q}\text{th}$ UT from the $k_{q}\text{th}$ UT}}\\ \rowcolor{lightblue}
			\hline
			$\varrho_{k_{q}}$ &\parbox{7cm}{\centering The channel temporal correlation function of the $k_q\text{th}$ UT}\\
			\hline
			$\mathbf{s}_q$ &\parbox{7cm}{\centering The basic pilot vector of the $q\text{th}$ group}\\ \rowcolor{lightblue}
			\hline
			$\mathbf{S}_q$ &\parbox{7cm}{\centering The basic pilot matrix of the $q\text{th}$ group generated by $\mathbf{s}_q$}\\
			\hline
			$\mathbf{X}_{k_q}$ &\parbox{7cm}{\centering The pilot of the $k_q\text{th}$ UT}\\ \rowcolor{lightblue}
			\hline
			$\varphi_{k_q}$ &\parbox{7cm}{\centering The phase shift factor of the $k_q\text{th}$ UT}\\
			\hline
			$\mathbf{D}_{\varphi _{k_q}}$ &\parbox{7cm}{\centering The phase shift matrix of the $k_q\text{th}$ UT}\\ \rowcolor{lightblue}
			\hline
			$p_\mathrm{xtr}$ & The pilot signal transmit power\\
			\hline
			$\eta_{\mathrm{tr}}$ & The signal-to-noise ratio over the $\ell\text{th}$ OFDM symbol\\
			\hline
			$\mathbf{Y}_\ell$ &\parbox{7cm}{\centering The space-frequency domain received signal at the $\ell\text{th}$ OFDM symbol}\\
			\hline
			$\mathbf{Y}_{k_{q},\ell}$ &\parbox{7cm}{\centering The LS estimate of $\mathbf{H}_{k_{q},\ell}$}\\ \rowcolor{lightblue}
			\hline
			$\boldsymbol{\chi}_{N_\mathrm{c},\phi}^\mathfrak{r}$ &\raisebox{-0.7ex}{\parbox{7cm}{\centering The Zadoff-Chu (ZC) sequence of length $N_\mathrm{c}$ with the root index of $\mathfrak{r}$ and a cyclic shift of $\phi$}}\\
			\hline
			$\mathbf{R}^{q}_{q'}$ &\parbox{7cm}{\centering The space-frequency domain cross-correlation matrix of basic pilot matrices $\mathbf{S}_{q}$ and $\mathbf{S}_{q'}$}\\ \rowcolor{lightblue}
			\hline
			$\mathbf{R}_{k^{'}_{q'}}^{k_{q}}$ &\raisebox{-0.6ex}{\parbox{7cm}{\centering The space-frequency domain normalized cross-correlation matrix of pilot $\mathbf{X}_{k_q}$ and $\mathbf{X}_{k'_{q'}}$}}\\
			\hline
			$\mathbf{Z}^{q}_{q'}$ &\raisebox{-0.5ex}{\parbox{7cm}{\centering The angle-delay domain cross-correlation matrix of basic pilot matrices $\mathbf{S}_{q}$ and $\mathbf{S}_{q'}$}}\\ \rowcolor{lightblue}
			\hline
			$\mathbf{Z}_{k'_{q'}}^{k_q}$ &\raisebox{-0.5ex}{\parbox{7cm}{\centering The angle-delay domain cross-correlation matrix of UT $k_q$ and $k'_{q'}$}}\\
			\hline
			$\Theta^{q'}_{q}$ &\parbox{7cm}{\centering The prior pilot argument information (PAI) of the $q'\text{th}$ group UT for the $q\text{th}$ group}\\ \rowcolor{lightblue}
			\bottomrule
		\end{tabular}
	\end{table}

	In this paper, we propose a method for channel acquisition in massive MIMO-OFDM systems utilizing multi-group adjustable phase shift pilots (MAPSPs). Prior studies \cite{YCLY2021,you2015channel,XNLY2020} employed single basic pilot matrix for APSP implementation. \cite{JTXG2024} extended APSP by introducing Doppler domain block sparsity beyond classical angle-delay domain sparsity, proposing triple-beam time-frequency shift pilots. This work advances the APSP framework from a novel perspective. The proposed MAPSP leverages both channel characteristics and pilot argument information (PAI) to categorize user terminals (UTs) into multiple groups. By properly scheduling MAPSPs and received signal pre-processing, pilot interference in the system can be mitigated, and the mean squared error (MSE) of the channel estimation can be minimized. In future mobile communication systems designed for high SE, MAPSP achieves robust channel acquisition performance without increasing pilot overhead by leveraging the structural characteristics of both channel and pilots.
	
	The main contributions are summarized as follows:
	
	\begin{itemize}
		
		\item We provide a massive MIMO-OFDM system model characterized by significant sparsity. Subsequently, we propose a MAPSP channel acquisition method by extending a single APSP group to multiple groups in TDD mode. Then, we conduct a  comprehensive analysis of the performance of both methods utilizing MMSE estimation.
		
		\item We establish a unified mathematical expression for inter/intra-group pilot pollution in MAPSP, developing a discrete Fourier transform (DFT)-based fast calculation method for pilot interference. Through a structural decomposition of the interference term matrices, we derive the optimal design conditions for the basic pilot matrices. Building on these findings, a novel MAPSP implementation scheme using Zadoff-Chu (ZC) sequences is proposed, which includes a received signal pre-processing method and a dual-layer (inter-group and intra-group) scheduling algorithm. The base station (BS) preprocesses uplink signals to extract intra-group channels, reducing the problem to single-group APSP estimation.
		
		\item The performance of the proposed MAPSP is verified across three typical scenarios in the 3GPP 38.901 protocol. Specifically, when the number of UTs is 84 at SNR = 30 dB, MAPSP achieves approximately $17.2\%$, $10.7\%$, and $8.5\%$ higher SE compared to APSP, respectively. When scaling to 126 UTs, these gains surge dramatically to $253.4\%$, $175.2\%$, and $117.5\%$. The simulation results not only indicate that MAPSP achieves significant improvements in SE, but also shows considerable potential for numerous UT channel estimations.
	\end{itemize}	
	
	\subsection{Notations}
	
	We use $\bar{\imath} = \sqrt{-1}$ to denote the imaginary unit. $\left\langle a\right\rangle_b$ denotes the operation of $a$ modulo $b$. Upper (lower) case boldface letters denote matrices (column vectors). The notation $\triangleq $ is used for definitions. Notation $\sim$ represents “distributed as”. We adopt $\mathbf{I}_{N}$ to denote the $N\times N$ identity matrix, and $\mathbf{I}_{N\times L}$ to denote the matrix composed of the first $L\ (\le N)$ columns of $\mathbf{I}_{N}$. We adopt $\mathbf{0}$ to denote the zero vector or matrix. We use superscripts $(\cdot)^T$, $(\cdot)^H$, and $(\cdot)^*$ to denote the transpose,  conjugate-transpose, and conjugate operations, respectively. The operator $\operatorname{diag}\{\mathbf{x}\}$ denotes the diagonal matrix with the components of $\mathbf{x}$ along its main diagonal. $[\cdot]$ with subscripts is employed to denote the elements of the vector or matrix, and $[\mathbf{x}]_i$, $[\mathbf{X}]_{i,j}$, $[\mathbf{X}]_{i,:}$, $[\mathbf{X}]_{:,j}$ denote the $i\text{th}$ element of the vector $\mathbf{x}$, the $(i,j)\text{th}$ element of the matrix $\mathbf{X}$, the $i\text{th}$ row and $j\text{th}$ column of the matrix $\mathbf{X}$, respectively, where the element indices start with $0$.  $\left\lfloor x\right\rfloor$ and $\left\lceil x\right\rceil$ denote the integer rounding down and up on $x$, respectively. $\mathbb{R}^{M\times N}$ and $\mathbb{C}^{M\times N}$ denote the spaces of $M\times N$ matrices of real and complex components, respectively. $\mathsf{E}\{\cdot\}$ denotes the expectation operation. $\mathcal{N}(a,B)$, $\mathcal{CN}(a,B)$, and $\mathcal{WN}(a,B)$ denote the Gaussian, circular symmetric complex Gaussian, and wrapped Gaussian distributions with mean $a$ and covariance $B$, respectively. $\odot$ denotes the Hadamard product. $\Re\left\{\cdot\right\}$ and $\Im\left\{\cdot\right\}$ denote the real and imaginary parts of a complex vector or matrix, respectively. $\mathbf{W}_{N}$, $\mathbf{W}_{N\times L}$, and $\mathbf{W}_{N,l}$ denote the $N$-dimensional DFT matrix, the matrix composed of the first $L\ (\le N)$ columns of $\mathbf{W}_{N}$, and the $l\text{th}$ column of $\mathbf{W}_{N}$, respectively. $\mathscr{F}\left\{\mathbf{x}\right\}$ and $\mathscr{F}^{-1}\left\{\mathbf{x}\right\}$ denote DFT and inverse DFT (IDFT) of $\mathbf{x}$, respectively. The notation $\backslash$ denotes the operation of set subtraction. We define the cyclic shift matrix,
	\begin{align}
		\mathbf{\Lambda }_N^n\triangleq
		\begin{bmatrix}\mathbf{0}&\mathbf{I}_{N-\left\langle n \right \rangle_N}\\\mathbf{I}_{\left\langle n \right \rangle_N}&\mathbf{0}
		\end{bmatrix}\nonumber.
	\end{align}
	For convenience, some important symbol definitions in this paper are summarized in \tabref{tb:notation}.
	
	\subsection{Outline}
	
	The rest of this paper is summarized as follows. In Section \ref{sec_model}, the system model is constructed. In Section \ref{sec_channel acquisition}, we propose an MMSE channel estimation method with MAPSP. In Section \ref{sec_best condition}, we analyze the components of the pilot interference. In Section \ref{sec_implementation}, we give an implementation scheme including pilot scheduling and pre-processing process to alleviate pilot interference. Section \ref{sec_simulation} and Section \ref{sec_conclusion} show the simulation results and conclusions, respectively.
	
	\section{Massive MIMO-OFDM System Model}\label{sec_model}
	
	\subsection{System Setup}
	
	We consider a single-cell TDD wideband massive MIMO wireless communication system. The central BS is equipped with a uniform linear array (ULA) with $M$ antennas, and each antenna is separated by one-half wavelength $\lambda$. The cell accommodates $K$ single-antenna UTs. To facilitate the description, we divide $K$ UTs into $Q$ groups in advance. We denote the set of group as $\mathcal{Q}=\{0,1,\cdots ,Q-1\}$, where $q\in\mathcal{Q}$ represents the group index. The $q\text{th}$ UT set is denoted as $\mathcal{K}_q=\{0,1,\cdots ,K_q-1\}$, where $\sum_{q=0}^{Q-1}K_q=K$ and $k_q\in\mathcal{K}_q$ represents the UT index in the $q\text{th}$ group. Please note, $q$ and $k_q$ in superscripts/subscripts denote the $q\mathrm{th}$ group and the $k_q\mathrm{th}$ UT in the $q\mathrm{th}$ group, respectively. We assume that the channels of different UTs are statistically independent. 
	
	We utilize OFDM modulation with $N_\mathrm{c}$ subcarriers, which can be achieved by $N_\mathrm{c}$-point IDFT. The length of the cyclic prefix (CP) is $N_\mathrm{g}\ (\le N_\mathrm{c})$. We denote $T_\mathrm{sym}=(N_\mathrm{c}+N_\mathrm{g})T_\mathrm{s}$ and $T_\mathrm{c}=N_\mathrm{c}T_\mathrm{s}$ as the durations of system sampling with and without CP, respectively, where $T_\mathrm{s}$ is the system sampling duration. It is assumed that the time utilized for the CP is greater than the maximum channel delay for all UTs.
	
	\subsection{Channel Model}
	
	We assume that the channel remains unchanged during one OFDM symbol and varies between symbols. We utilize the physically motivated massive MIMO-OFDM channel model to formulate the problem. In the UL, we set the channel response vector between the $k_q\text{th}$ UT and the $m\text{th}$ antenna of the BS over the $\ell\text{th}$ OFDM symbol at the $n_c\text{th}$ subcarrier is represented as
	\begin{align}
		\left[\mathbf{g}_{k_q,\ell}^{n_c}\right]_m=&\sum_{p=0}^{N_{\mathrm{p}}-1}\gamma_{k_q,\ell}^{p}\exp(-\bar{\imath}\pi m\cos (a_{k_q,p}))\nonumber\\
		&\cdot\exp\left(-\bar{\imath}2\pi n_c\frac{\tau_{k_q,p}}{N_\mathrm{c}T_\mathrm{s}}\right),
	\end{align}
	where $\mathbf{g}_{k_q,\ell}^{n_c}\in\mathbb{C}^{M\times 1}$ and $N_{\mathrm{p}}$ is the total number of paths. $\gamma_{k_q,\ell}^{p}$, $a_{k_q,p}$, $\tau_{k_q,p}$ represent complex gain, directional cosine, and delay, respectively. In particular, the complex gain of the LoS path is denoted as
	\begin{equation}
		\gamma_{k_q,\ell}^{0}=\beta_{k_q,\ell}^{0}\exp(-\frac{\bar{\imath}2\pi}{\lambda}d_0),
	\end{equation}   
	where $\beta_{k_q,\ell}^{0}$ is the real gain and $d_0$ is the distance from UT to BS \cite{MGNZ2023}. We define the exponent of complex gain as the argument. Taking the LoS path as an example, the argument is $-\frac{2\pi}{\lambda}d_0$. For NLoS paths gain $\gamma_{k_q,\ell}^{p}\ (p>0)$, the real gain is relatively small and the argument is random, compared to LoS path \cite{THDC2018}. Because of the significant power disparities among paths, the real gain and argument of $[\mathbf{g}_{k_q,\ell}^{n_c}]_m$ are primarily influenced by the LoS path. 
	
	By gathering the channel response vectors of different subcarriers together, the space-frequency domain channel response matrix $\mathbf{G}_{k_q,\ell}$ is obtained as follows.
	\begin{equation}
		\mathbf{G}_{k_q,\ell}=\left[\mathbf{g}_{k_q,\ell}^{0},\mathbf{g}_{k_q,\ell}^{1},\dots,\mathbf{g}_{k_q,\ell}^{N_\mathrm{c}-1}\right]\in\mathbb{C}^{M\times N_\mathrm{c}}.
	\end{equation}
	In massive MIMO-OFDM systems, calculating the high-dimensional matrix $\mathbf{G}_{k_q,\ell}$ poses significant challenges. For simplicity, we estimate the channel in the angle-delay domain through the eigenvalue decomposition of $\mathbf{G}_{k_q,\ell}$ \cite{you2015channel}. 
	\begin{equation}\label{Channel unitary equivalence}
		\begin{aligned}
			\left[\mathbf{G}_{k_q,\ell}\right]_{i,j}
			=&\frac{1}{\sqrt{N_\mathrm{c}}}\left[\mathbf{A}\mathbf{H}_{k_q,\ell}\mathbf{W}_{N_\mathrm{c}\times N_\mathrm{g}}^T\right]_{i,j}\\
			=&\frac{1}{\sqrt{N_\mathrm{c}}}\sum_{n=0}^{N_\mathrm{g}-1}(\left[\mathbf{A}\right]_{i,:}\mathbf{h}_{k_q,\ell}^{n})\left[\mathbf{W}_{N_\mathrm{c}\times N_\mathrm{g}}^T\right]_{n,j},
		\end{aligned}
	\end{equation}
	where 
	\begin{equation}
		\mathbf{H}_{k_q,\ell}=\left(\mathbf{h}_{k_q,\ell}^{0},\mathbf{h}_{k_q,\ell}^{1},\cdots,\mathbf{h}_{k_q,\ell}^{N_{\mathrm{g}}-1}\right)\in \mathbb{C}^{M\times N_\mathrm{g}},
	\end{equation} 
	is the angle-delay domain channel response matrix of the $k_q\text{th}$ UT at the $\ell$ symbol. The array response vector $\mathbf{A}\in \mathbb{C}^{M\times M}$ is detailed as
	\begin{equation}
		[\mathbf{A}]_{i,j}\triangleq\frac{1}{\sqrt{M}}\exp\left(-\bar{\imath}2\pi\frac{i(j-M/2)}{M}\right).
	\end{equation} 
	
	By utilizing (\ref{Channel unitary equivalence}), the estimated channel can be converted to $\mathbf{H}_{k_q,\ell}$. So, we focus on the characteristics of the angle-delay domain channel. The element $[\mathbf{H}_{k_q,\ell}]_{i,j}$ is denoted as
	\begin{align}\label{h ele model}
		\left[\mathbf{h}_{k_q,\ell}^{j}\right]_{i}=&\frac{1}{\sqrt{N_\mathrm{c}}}\sum_{n=0}^{N_\mathrm{c}-1}(\left[\mathbf{A}^H\right]_{i,:}\mathbf{g}_{k_q,\ell}^{n})\left[\mathbf{W}_{N_\mathrm{c}\times N_\mathrm{g}}^*\right]_{n,j}\nonumber\\
		=&\frac{1}{\sqrt{N_\mathrm{c}}}\sum_{n=0}^{N_\mathrm{c}-1}\left[\mathbf{W}_{N_\mathrm{c}\times N_\mathrm{g}}^*\right]_{n,j}\sum_{m=0}^{M-1}\left[\mathbf{A}^H\right]_{i,m}\left[\mathbf{g}_{k_q,\ell}^{n}\right]_{m}.
	\end{align}
	Assuming that the elements of $\mathbf{G}_{k_q,\ell}$ are independent, then the transformation from $\mathbf{g}_{k_q,\ell}^{n}$ to $\mathbf{h}_{k_q,\ell}^{j}$ can be computed element-wise in both the spatial and frequency domains as (\ref{h ele model}). We consider a mobility scenario with a large number of UTs, where frequent UT movement leads to variations in the communication link state. Since the LoS component dominates channel variations, we assume that the phases of the NLoS components resemble those of the LoS component. Under this assumption, the probability distribution of the received signal phase within $[0,2\pi)$ appears to be unimodal. To approximate this phase distribution, we employ the wrapped Gaussian distribution $\mathcal{WN}(\bar{\mu}_{\mathrm{w}},\bar{\sigma}_{\mathrm{w}}^2)$, where the mean $\bar{\mu}_{\mathrm{w}}$ is related to the argument of the LoS component, and the variance $\bar{\sigma}_{\mathrm{w}}^2$ reflects the similarity between the arguments of the NLoS and LoS components. Rewrite (\ref{h ele model}) as follows
	\begin{equation}\label{h model}
		\left[\mathbf{h}_{k_q,\ell}^{j}\right]_{i}=\left[\boldsymbol{\alpha}_{k_q,\ell}^{j}\right]_{i}\exp(\bar{\imath}\bar{\theta}_{i,j}^{k_q}),
	\end{equation}
	where $\boldsymbol{\alpha}_{k_q,\ell}^{j}\in \mathbb{C}^{M \times 1}$ represents the real gain in angle-delay domain, and argument $\bar{\theta}_{i,j}^{k_q}\sim\mathcal{WN}(\bar{\mu}_{i,j}^{k_q},\bar{\sigma}^2)$. Specifically, the phase is predominantly clustered around a particular mean value, exhibiting a variance that is significantly smaller than $2\pi$ \cite{JWSL2023}. In this case, this wrapped Gaussian distribution can be well approximated by Gaussian distribution $\mathcal{N}(\bar{\mu}_{i,j}^{k_q},\bar{\sigma}^2)$.
	
	The matrix $\mathbf{H}_{k_q,\ell}$ is related to the received signal delay and the angle of arrival (AOA). Previous research has shown the sparsity of massive MIMO-OFDM channels in the angle-delay domain \cite{you2015channel}. Based on the statistical characteristics of $\mathbf{H}_{k_q,\ell}$, the relationship between it and the angle-delay domain channel power matrix $\mathbf{P}_{k_q}$ can be established when the number of antennas is sufficiently large \cite{you2015channel}. 
	\begin{equation}\label{Pilot channel power}
		\mathsf{E}\left\{\mathbf{H}_{k_q,\ell+\Delta_{\ell}}\odot\mathbf{H}_{k_q,\ell}^*\right\}=\varrho_{k_{q}}(\Delta_{\ell})\mathbf{P}_{k_q},
	\end{equation}
	\begin{equation}
		\boldsymbol{\alpha}_{k_q,\ell}^{n_g}=\sqrt{\left[\mathbf{P}_{k_q}\right]_{:,n_g}},
	\end{equation}
	where $\varrho_{k_{q}}(\Delta_{\ell})$ is the channel temporal correlation function (TCF) and related to the Doppler frequency parameter $\nu$, $T_{\mathrm{sym}}$, and $\Delta_\ell$. Concerning the Clarke-Jakes channel power Doppler spectrum model, the corresponding TCF can be succinctly represented by the zeroth-order Bessel function of the first kind $\mathrm{J}_{0}(\cdot)$ \cite{MP2011, WJDC1994}.
	\begin{equation}
		\varrho_{k_{q}}(\Delta_\ell)=\mathrm{J}_0(2\pi\nu T_\mathrm{sym}\Delta_\ell).
	\end{equation}
	Using $\mathrm{J}_{0}(\cdot)$ naturally satisfies $\varrho_{k_{q}}(0)=1$ and inherently models $\varrho_{k_{q}}(\Delta_\ell)=\varrho_{k_{q}}(-\Delta_\ell)$ via its even symmetry. For massive MIMO-OFDM channels, different elements of $\mathbf{H}_{k_q,\ell}$ are mutually statistically uncorrelated \cite{YCLY2021J}. Therefore, the corresponding $\mathbf{P}_{k_q}$ exhibits sparsity in the angle-delay domain. Since $\mathbf{P}_{k_q}$ is composed of variances of independent elements, it can be estimated element by element \cite{YCLY2021,you2015channel,SDSLF2023}. Notably, the channel power matrix represents a form of slowly-varying statistical CSI, and various established methods exist for acquiring $\mathbf{P}_{k_q}$ \cite{CSXG2015, CWSJ2015, JTLY2025}.
	
	In addition to the sparsity in the angle-delay domain, our model further captures the non-uniformity of the argument. Building upon this characteristic, multiple UTs with overlapping angle-delay domain channels can be effectively clustered into distinct groups, where the arguments within each group are mutually offset. 
	
	For the rest of the paper, we assume that all the power matrices $\mathbf{P}_{k_q}$ of all UTs and argument distribution are known by the BS. 
	
	\section{Channel Acquisition With MAPSPs}\label{sec_channel acquisition}
	
	Based on the proposed massive MIMO-OFDM channel model, we extend APSP in \cite{you2015channel} into multi-groups for channel acquisition to accommodate more UTs. In this section, we will present the pilot structure and the method of channel acquisition with MAPSPs. We also prove that intra-group interference is a special case of inter-group interference, which means APSP is a special case of MAPSP.
	
	\subsection{Structure of MAPSP}
	
	We assume that the BS uses the time advance (TA) mechanism to ensure all UTs are completely synchronized \cite{SDSLF2023}. On the $\ell\text{th}$ OFDM symbol of each frame during the UL, all UTs transmit their pilots simultaneously, and the BS receives all pilot signals. As mentioned earlier, the UTs are divided into $Q$ groups, each of which is assigned a basic pilot matrix. We design each group of UTs to generate APSP using one basic pilot matrix that differs between groups. The previous channel acquisition method with APSP can be regarded as a special case of MAPSP when $Q=1$. The pilot in the frequency domain for UT $k_q$ in group $q$ is given as
	\begin{equation}
		\mathbf{X}_{k_q}\triangleq\sqrt{p_{\mathrm{xtr}}}\mathbf{D}_{\varphi _{k_q}}\mathbf{S}_q\in \mathbb{C}^{N_\mathrm{c}\times N_\mathrm{c}},
	\end{equation}
	where $\varphi_{k_q}=0,1,\ldots,N_{\mathrm{c}}-1,\ \mathbf{D}_{\varphi _{k_q}}\triangleq\operatorname{diag}\left\{\mathbf{W}_{N_{\mathrm{c}},\varphi _{k_q}}\right\}$ is the phase shift factor of the $k_q\text{th}$ UT, $p_\mathrm{xtr}$ is the pilot signal transmit power and $\mathbf{S}_q\triangleq\operatorname{diag}\left\{\mathbf{s}_q\right\}\in \mathbb{C}^{N_\mathrm{c}\times N_\mathrm{c}}$ is the basic pilot matrix of group $q$ which satisfies $\mathbf{S}_q\mathbf{S}_q^H=\mathbf{I}_{N_\mathrm{c}}$. 
	
	The space-frequency domain signal received by BS at the $\ell\text{th}$ OFDM symbol is represented as
	\begin{equation}\label{received signal}
		\mathbf{Y}_\ell=\sum_{q=0}^{Q-1}\sum_{k_{q}=0}^{K_{q}-1}\mathbf{G}_{k_{q},\ell}\mathbf{X}_{k_{q}}+\mathbf{N}_\ell,
	\end{equation}
	where $\mathbf{Y}_\ell\in\mathbb{C}^{M\times N_\mathrm{c}}$ represents the received signal, $\mathbf{G}_{k_{q},\ell}$ is the space-frequency domain channel response matrix, and $\mathbf{N}_\ell$ is the additive white Gaussian noise (AWGN) matrix with identical and independent distributed (i.i.d.) elements. During the UL pilot segment, we assume that the AWGN $\mathbf{N}_\ell\sim\mathcal{CN}(0,p_{\mathrm{ntr}})$ and $p_{\mathrm{ntr}}$ represents the noise power. 
	
	The received signal $\mathbf{Y}_\ell$ of BS includes the space-frequency domain channel and pilot information of all UTs, and $[\mathbf{Y}_{\ell}]_{i,j}$ denotes the received signal over the $j\text{th}$ subcarrier at the $i\text{th}$ antenna. To facilitate the subsequent $\mathbf{Y}_\ell$ decorrelation and channel estimation expressions, the expression methods of autocorrelation and cross-correlation of the pilot matrix are given here in advance:
	\begin{equation}
		\begin{aligned}
			\mathbf{X}_{k_q}\mathbf{X}_{k'_{q'}}^H&=p_{\mathrm{xtr}}\mathbf{D}_{\varphi _{k_q}}\mathbf{S}_q\mathbf{S}_{q'}^H\mathbf{D}_{\varphi _{k'_{q'}}}^H\\
			&=p_{\mathrm{xtr}}\mathbf{D}_{\varphi _{k_q}-\varphi _{k'_{q'}}}\mathbf{R}^{q}_{q'}=p_{\mathrm{xtr}}\mathbf{R}^{k_q}_{k'_{q'}}\label{Diff Pilot cross-correlation}.
		\end{aligned}
	\end{equation}
	Because the basic pilot matrix $\mathbf{S}_q$ is different, the cross-correlation matrix expression of the pilot matrices for different UT groups has great uncertainty, denoted as space-frequency domain pilot cross-correlation matrix (SFPCM), $\mathbf{R}^{k_q}_{k'_{q'}}=\mathbf{D}_{\varphi _{k_q}}\mathbf{S}_q\mathbf{S}_{q'}^H\mathbf{D}_{\varphi _{k'_{q'}}}^H$. Matrix $\mathbf{R}_{q'}^{q}\triangleq\mathbf{S}_q\mathbf{S}_{q'}^H$ is the cross-correlation of basic pilot matrices. Observe that $\mathbf{R}_{q'}^{q}=\mathbf{S}_q\mathbf{S}_{q'}^H$ and $\mathbf{R}_{q}^{q'}\triangleq\mathbf{S}_{q'}\mathbf{S}_{q}^H$ are distinct. Note that the correlation of the pilot matrix can be divided into three kinds, 
	\begin{itemize}
		\item Autocorrelation, $\mathbf{R}_{k_q}^{k_{q}}=\mathbf{I}_{N_\mathrm{c}}$.
		\item Cross-correlation of the same group, $\mathbf{R}^{k_q}_{k'_{q}}=\mathbf{D}_{\varphi _{k_q}-\varphi _{k'_q}}$.
		\item Cross-correlation of different groups, (\ref{Diff Pilot cross-correlation}).
	\end{itemize}
	We will examine $\mathbf{R}^{k_q}_{k'_{q'}}$ comprehensively in Section \ref{sec_best condition}. 
	
	\subsection{Channel Estimation}
	
	If the MMSE estimation is used directly to obtain the channel information $\mathbf{G}_{k_{q},\ell}$, the computational complexity is enormous. Fortunately, we can use the channel sparsity mentioned above to estimate $\hat{\mathbf{H}}_{k_q,\ell}$ in the angle-delay domain, and then use the unit-equivalent relation in (\ref{Channel unitary equivalence}) to obtain $\hat{\mathbf{G}}_{k_{q},\ell}$ indirectly. The received signal (\ref{received signal}) on the BS side of the UL can be rewritten as
	\begin{align}
		\mathbf{Y}_\ell=\frac{1}{\sqrt{N_\mathrm{c}}}\cdot\sum_{q=0}^{Q-1}\sum_{k_{q}=0}^{K_{q}-1}\mathbf{A}\mathbf{H}_{k_q,\ell}\mathbf{W}_{N_c\times N_g}^T\mathbf{X}_{k_q}+\mathbf{N}_\ell.
	\end{align}
	We assume to estimate the channel information of the $k'_{q'}\text{th}$ UT. After decorrelation and power normalization of $\mathbf{Y}_\ell$, the least squares (LS) estimate of $\mathbf{H}_{k'_{q'},\ell}$ is obtained as follows
	\begin{align}\label{LS}
		\mathbf{Y}_{k'_{q'},\ell} =&\frac{1}{p_{\mathrm{xtr}}\sqrt{N_{\mathrm{c}}}}\mathbf{A}^{H}\mathbf{Y}_{\ell}\mathbf{X}_{k'_{q'}}^{H}\mathbf{W}_{N_{\mathrm{c}}\times N_{\mathrm{g}}}^{*} \nonumber\\
		=&\underbrace{\frac{1}{N_{\mathrm{c}}}\sum_{k_{q'}\in  \mathcal{K}_{q'}\backslash k'_{q'}}\mathbf{H}_{k_{q'},\ell}\mathbf{W}_{N_{\mathrm{c}}\times N_{\mathrm{g}}}^T\mathbf{D}_{\varphi _{k_{q'}}-\varphi _{k'_{q'}}}\mathbf{W}_{N_{\mathrm{c}}\times N_{\mathrm{g}}}^*}_{\text{Intra-group pilot interference}} \nonumber\\
		&+\underbrace{\frac{1}{N_{\mathrm{c}}}\sum_{q\in\mathcal{Q}\backslash q'}\sum_{k_q=0}^{K_q-1}\mathbf{H}_{k_q,\ell}\mathbf{W}_{N_{\mathrm{c}}\times N_{\mathrm{g}}}^T\mathbf{R}_{k'_{q'}}^{k_q}\mathbf{W}_{N_{\mathrm{c}}\times N_{\mathrm{g}}}^*}_{\text{Inter-group pilot interference}} \nonumber\\
		&+\mathbf{H}_{k'_{q'},\ell}+\underbrace{\frac{1}{p_{\mathrm{xtr}}\sqrt{N_{\mathrm{c}}}}\mathbf{A}^{H}\mathbf{N}_{\ell}\mathbf{X}_{k'_{q'}}^{H}\mathbf{W}_{N_{\mathrm{c}}\times N_{\mathrm{g}}}^{*}}_{\text{Channel noise}},	
	\end{align}
	where the third term $\mathbf{H}_{k'_{q'},\ell}$ in the right hand side is the angle-delay domain channel response of the UT $k'_{q'}$. 
	
	In (\ref{LS}), the first term in the right hand side denotes the pilot interference from the same group of UTs. We can simplify it to the form of cyclic shift of $\bar{\mathbf{H}}_{k_{q'},\ell}$ \cite{you2015channel},
	\begin{align}\label{same group interference}
			\mathbf{H}_{k_{q'},\ell}^{\varphi_{k_{q'}}-\varphi_{k'_{q'}}}& =\mathbf{H}_{k_{q'},\ell}\mathbf{W}_{N_{\mathrm{c}}\times N_{\mathrm{g}}}^T\mathbf{D}_{\varphi_{k_{q'}}-\varphi_{k'_{q'}}}\mathbf{W}_{N_{\mathrm{c}}\times N_{\mathrm{g}}}^* \nonumber\\
			&=\bar{\mathbf{H}}_{k_{q'},\ell}\mathbf{W}_{N_{\mathrm{c}}}^{T}\mathbf{D}_{\varphi_{k_{q'}}-\varphi_{k'_{q'}}}\mathbf{W}_{N_{\mathrm{c}}}^{*}\mathbf{I}_{N_{\mathrm{c}}\times N_{\mathrm{g}}} \nonumber\\
			&=N_{\mathrm{c}}\cdot\bar{\mathbf{H}}_{k_{q'},\ell}\mathbf{\Lambda }_{N_{\mathrm{c}}}^{\varphi_{k_{q'}}-\varphi_{k'_{q'}}}\mathbf{I}_{N_{\mathrm{c}}\times N_{\mathrm{g}}},
	\end{align}
	where
	\begin{equation}
		\begin{aligned}
			\bar{\mathbf{H}}_{k_{q'},\ell}& \triangleq\mathbf{H}_{k_{q'},\ell}\mathbf{I}_{N_{\mathrm{c}}\times N_{\mathrm{g}}}^{T} =[\mathbf{H}_{k_{q'},\ell},\ \mathbf{0}_{M\times(N_{\mathrm{c}}-N_{\mathrm{g}})}],
		\end{aligned}
	\end{equation}
	is a complement-0 extension of $\mathbf{H}_{k_{q'},\ell}$. The pilot interference term from the same UT group $k_{q'}$ can be understood as the result of a cyclic shift of $\bar{\mathbf{H}}_{k_{q'},\ell}$ to the right by $\varphi_{k_{q'}}-\varphi_{k'_{q'}}$ units before intercepting the first $N_{\mathrm{g}}$ columns.
	
	The pilot interference from different groups of UTs is the second term in the right hand side in (\ref{LS}). The specific expression of this term varies with the basic pilot matrix, and we give a general representation as follows
	\begin{equation}\label{diff group interference}
		\begin{aligned}
			\mathbf{H}_{k_q,\ell}^{\varphi_{k_{q}}-\varphi_{k'_{q'}}}& =\mathbf{H}_{k_q,\ell}\mathbf{W}_{N_{\mathrm{c}}\times N_{\mathrm{g}}}^T\mathbf{R}_{k'_{q'}}^{k_q}\mathbf{W}_{N_{\mathrm{c}}\times N_{\mathrm{g}}}^* \\
			&=\mathbf{H}_{k_q,\ell}\mathbf{I}_{N_{\mathrm{c}}\times N_{\mathrm{g}}}^T\mathbf{W}_{N_{\mathrm{c}}}^T\mathbf{R}_{k'_{q'}}^{k_q}\mathbf{W}_{N_{\mathrm{c}}}^*\mathbf{I}_{N_{\mathrm{c}}\times N_{\mathrm{g}}} \\
			&=N_{\mathrm{c}}\cdot\bar{\mathbf{H}}_{k_q,\ell}\mathbf{Z}_{k'_{q'}}^{k_q}\mathbf{I}_{N_{\mathrm{c}}\times N_{\mathrm{g}}},
		\end{aligned}
	\end{equation}
	where
	\begin{equation}\label{ADPCM}
		\mathbf{Z}_{k'_{q'}}^{k_q}\triangleq\frac{1}{N_{\mathrm{c}}}\mathbf{W}_{N_{\mathrm{c}}}^T\mathbf{R}_{k'_{q'}}^{k_q}\mathbf{W}_{N_{\mathrm{c}}}^*,
	\end{equation}
	is defined as the angle-delay domain pilot cross-correlation matrix (ADPCM). The pilot interference term from the other UT groups $k_{q}$ can be viewed as the result of multiplying $\bar{\mathbf{H}}_{k_{q},\ell}$ and ADPCM before intercepting the first $N_{\mathrm{g}}$ columns. Pilot interference from the same group of UT can be considered a special case of (\ref{ADPCM}), when 
	\begin{equation}
		\mathbf{Z}_{k'_{q'}}^{k_{q'}}=\mathbf{\Lambda }_{N_{\mathrm{c}}}^{\varphi_{k_{q'}}-\varphi_{k'_{q'}}}.
	\end{equation}
	It is worth noting that the pilot noise, the fourth term in the right hand side in (\ref{LS}), can be proved to obey a circular symmetric complex Gaussian distribution $\mathcal{CN}(0,p_{\mathrm{ntr}}/p_{\mathrm{xtr}})$ by using the unitary transformation property, expressed as
	\begin{equation}\label{Gaussian noise}
		\frac{1}{\sqrt{\eta_{\mathrm{tr}} N_{\mathrm{c}}}}\mathbf{N}_{\mathrm{nor}}\triangleq \frac{1}{p_{\mathrm{xtr}}\sqrt{N_{\mathrm{c}}}}\mathbf{A}^{H}\mathbf{N}_{\ell}\mathbf{X}_{k'_{q'}}^{H}\mathbf{W}_{N_{\mathrm{c}}\times N_{\mathrm{g}}}^{*},
	\end{equation}
	where $\eta_{\mathrm{tr}}\triangleq p_{\mathrm{xtr}}/p_{\mathrm{ntr}}$ denotes the signal-to-noise ratio (SNR) over the $\ell\text{th}$ OFDM symbol, and $\mathbf{N}_{\mathrm{nor}}$ is the normalized AWGN matrix with each element being independent and obeying the same distribution $\mathcal{CN}(0,1)$.
	
	The received pilot signal after decorrelating includes not only $\mathbf{H}_{k'_{q'},\ell}$ to be estimated but also two types of pilot interference and AWGN. Substituting (\ref{same group interference}), (\ref{diff group interference}) and (\ref{Gaussian noise}) into (\ref{LS}), we have
	\begin{align}
		\mathbf{Y}_{k'_{q'},\ell}=&\mathbf{H}_{k'_{q'},\ell}+\frac{1}{N_{\mathrm{c}}}\sum_{k_{q'}\in  \mathcal{K}_{q'}\backslash k'_{q'}}\mathbf{H}_{k_{q'},\ell}^{\varphi_{k_{q'}}-\varphi_{k'_{q'}}}\nonumber\\
		&+\frac{1}{N_{\mathrm{c}}}\sum_{q\in\mathcal{Q}\backslash q'}\sum_{k_q=0}^{K_q-1}\mathbf{H}_{k_q,\ell}^{\varphi_{k_{q}}-\varphi_{k'_{q'}}}+\frac{1}{\sqrt{\eta_{\mathrm{tr}} N_{\mathrm{c}}}}\mathbf{N}_{\mathrm{nor}}\nonumber\\
		=&\frac{1}{N_{\mathrm{c}}}\sum_{q=0}^{Q-1}\sum_{k_q=0}^{K_q-1}\mathbf{H}_{k_q,\ell}^{\varphi_{k_{q}}-\varphi_{k'_{q'}}}+\frac{1}{\sqrt{\eta_{\mathrm{tr}} N_{\mathrm{c}}}}\mathbf{N}_{\mathrm{nor}}\nonumber\\
		=&\mathbf{H}_{\Sigma }+\frac{1}{\sqrt{\eta_{\mathrm{tr}} N_{\mathrm{c}}}}\mathbf{N}_{\mathrm{nor}},
	\end{align}
	where 
	\begin{equation}
		\mathbf{H}_{\Sigma }\triangleq\frac{1}{N_{\mathrm{c}}}\sum_{q=0}^{Q-1}\sum_{k_q=0}^{K_q-1}\mathbf{H}_{k_q,\ell}^{\varphi_{k_{q}}-\varphi_{k'_{q'}}}.
	\end{equation}
	Recalling (\ref{Pilot channel power}), since the elements of $\mathbf{H}_{k_{q'},\ell}$ are statistically uncorrelated, the elements of the pilot interference terms $\mathbf{H}_{k_{q'},\ell}^{\varphi_{k_{q'}}-\varphi_{k'_{q'}}}$ and $\mathbf{H}_{k_q,\ell}^{\varphi_{k_{q}}-\varphi_{k'_{q'}}}$ are also statistically uncorrelated. The power matrices of pilot interference from the same group and different groups are, respectively, defined as follows,
	\begin{equation}
		\begin{aligned}
			\mathbf{P}_{k_{q'}}^{\varphi_{k_{q'}}-\varphi_{k'_{q'}}} \triangleq&\frac{1}{N_{\mathrm{c}}^2}\mathsf{E}\left\{\mathbf{H}_{k_{q'},\ell}^{\varphi_{k_{q'}}-\varphi_{k'_{q'}}}\odot\left(\mathbf{H}_{k_{q'},\ell}^{\varphi_{k_{q'}}-\varphi_{k'_{q'}}}\right)^{*}\right\} \\
			=&\bar{\mathbf{P}}_{k_{q'}}\mathbf{\Lambda}_{N_{\mathrm{c}}}^{\varphi_{k_{q'}}-\varphi_{k'_{q'}}}\mathbf{I}_{N_{\mathrm{c}}\times N_{\mathrm{g}}},
		\end{aligned}
	\end{equation}
	\begin{equation}
		\mathbf{P}_{k_{q}}^{\varphi_{k_{q}}-\varphi_{k'_{q'}}}\triangleq\frac{1}{N_{\mathrm{c}}^2}\mathsf{E}\left\{\mathbf{H}_{k_{q},\ell}^{\varphi_{k_{q}}-\varphi_{k'_{q'}}}\odot\left(\mathbf{H}_{k_{q},\ell}^{\varphi_{k_{q}}-\varphi_{k'_{q'}}}\right)^{*}\right\},
	\end{equation}
	where
	\begin{equation}
		\begin{aligned}
			\bar{\mathbf{P}}_{k_{q'}}& \triangleq\mathbf{P}_{k_{q'}}\mathbf{I}_{N_{\mathrm{c}}\times N_{\mathrm{g}}}^{T}=[\mathbf{P}_{k_{q'}},\ \mathbf{0}_{M\times(N_{\mathrm{c}}-N_{\mathrm{g}})}]
		\end{aligned}
	\end{equation}
	is a complement-0 extension of $\mathbf{P}_{k_{q'}}$.
	
	
	Using the received signal $\mathbf{Y}_{k'_{q'},\ell}$ after decorrelating in (\ref{LS}), MMSE estimate $\hat{\mathbf{H}}_{k'_{q'},\ell}$ can be obtained element-by-element,
	\begin{equation}
		\begin{aligned}
			[\hat{\mathbf{H}}_{k'_{q'},\ell}]_{i,j}=&\frac{[\mathbf{P}_{k'_{q'}}]_{i,j}}{\sum_{q=0}^{Q-1}\sum_{k_q=0}^{K_q-1}\left[\mathbf{P}_{k_q}^{\varphi_{k_q}-\phi_{k'_{q'}}}\right]_{i,j}+\frac{1}{\eta _{\mathrm{tr}}}}[\mathbf{Y}_{k'_{q'},\ell}]_{i,j}\\
			=&\frac{[\mathbf{P}_{k'_{q'}}]_{i,j}}{\left[\mathbf{P}_\Sigma\right]_{i,j}+\frac{1}{\eta _{\mathrm{tr}}}}[\mathbf{Y}_{k'_{q'},\ell}]_{i,j},
		\end{aligned}
	\end{equation}
	where
	\begin{equation}
		\mathbf{P}_\Sigma\triangleq\sum_{q=0}^{Q-1}\sum_{k_q=0}^{K_q-1}\mathbf{P}_{k_q}^{\varphi_{k_q}-\phi_{k'_{q'}}}.
	\end{equation}
	Denote $\tilde{\mathbf{H}}_{k'_{q'},\ell} = \mathbf{H}_{k'_{q'},\ell}-\hat{\mathbf{H}}_{k'_{q'},\ell}$ as the channel estimation error of UT $k'_{q'}$, and we have
	\begin{align}\label{MMSE err}
			\sigma _{k'_{q'}}& \triangleq\sum_{i=0}^{M-1}\sum_{j=0}^{N_{\mathrm{g}}-1}\mathsf{E}\left\{\left|\left[\tilde{\mathbf{H}}_{k'_{q'},\ell}\right]_{i,j}\right|^{2}\right\}\nonumber \\
			&=\sum_{i=0}^{M-1}\sum_{j=0}^{N_{\mathrm{g}}-1}\mathsf{E}\left\{\left|\left[\mathbf{H}_{k'_{q'},\ell}\right]_{i,j}\right|^{2}-\left|\left[\hat{\mathbf{H}}_{k'_{q'},\ell}\right]_{i,j}\right|^{2}\right\}\nonumber \\
			&=\sum_{i=0}^{M-1}\sum_{j=0}^{N_{\mathrm{g}}-1}\left\{\left[\mathbf{P}_{k'_{q'}}\right]_{i,j}-\frac{\left[\mathbf{P}_{k'_{q'}}\right]_{i,j}^2}{\left[\mathbf{P}_\Sigma\right]_{i,j}+\frac{1}{\eta _\mathrm{tr}}}\right\}.
	\end{align}
	Obviously, the overall error of MMSE estimates for $K$ UTs can be calculated by summing the individual errors,
	\begin{equation}
		\sigma \triangleq\sum_{q=0}^{Q-1}\sum_{k_q=0}^{K_q-1}\sigma _{k_q}.
	\end{equation}
	The accuracy of channel estimation is directly affected by the intensity of pilot interference. By utilizing a pilot scheduling method, it is possible to minimize or eliminate pilot interference and approach the lower bound of MMSE channel estimation error in optimal conditions, denoted as
	\begin{equation}
		\begin{aligned}
			\sigma\geq&\sigma^{\mathrm{min}}\\
			=&\sum_{q=0}^{Q-1}\sum_{k_q=0}^{K_q-1}\sum_{i=0}^{M-1}\sum_{j=0}^{N_{\mathrm{g}}-1}\left\{\left[\mathbf{P}_{k_q}\right]_{i,j}-\frac{\left[\mathbf{P}_{k_q}\right]_{i,j}^{2}}{\left[\mathbf{P}_{k_q}\right]_{i,j}+\frac{1}{\eta _{\mathrm{tr}}}}\right\}.
		\end{aligned}
	\end{equation}
	The above formula can be considered as instance where all pilot interference has been eliminated. Specifically, for the $k'_{q'}\text{th}$ UT, $\mathbf{P}_\Sigma=\mathbf{P}_{k'_{q'}}$.

	\section{Analysis Of Pilot Interference}\label{sec_best condition}
	
	In this section, we conduct an analysis of the factors contributing to intra-group and inter-group pilot interference. We leverage the relationship between SFPCM and ADPCM to develop an efficient calculation method for pilot interference. Subsequently, we present two propositions that encapsulate the general expression for pilot interference and the conditions that the optimal basic pilot matrices must satisfy. 
	
	From a mathematical perspective, the sparsity of $\mathbf{H}_{k_q,\ell}$ can be explained in detail that only $c\ (\le N_{\mathrm{g}})$ column elements in matrix $\mathbf{H}_{k_q,\ell}$ are non-zero-valid, while the other elements are approximately zero. These non-zero column vectors are also only partially non-zero, which intuitively reflects the sparsity of the channel. In cases where UTs are utilizing the same basic pilot matrix, the channel overlap can be reduced by adjusting the phase shift of the pilots during scheduling \cite{you2015channel}. If $\forall k_{q'},k'_{q'}\in \mathcal{K}_q $ and $k_{q'}\neq k'_{q'}$,
	\begin{equation}
		\left(\bar{\mathbf{H}}_{k'_{q'},\ell}\mathbf{\Lambda }_{N_{\mathrm{c}}}^{\varphi_{k'_{q'}}}\right)\odot\left(\bar{\mathbf{H}}_{k_{q'},\ell}\mathbf{\Lambda }_{N_{\mathrm{c}}}^{\varphi_{k_{q'}}}\right)=\mathbf{0},
	\end{equation}
	then the pilot interference can be eliminated, thus reaching the lower bound. When the sparsity of the angle-delay domain is fully utilized, it means that the maximum UT number $\left\lfloor N_{\mathrm{c}}/c\right\rfloor $ can be achieved.
	 
	When MAPSPs are employed, pilot interference is categorized into two types: interference from the same group, i.e., intra-group interference, and different groups, i.e., inter-group interference. The intra-group interference is similar to APSPs generated from the single basic pilot matrix, but the inter-group interference is primarily influenced by ADPCM. We make the assumption about the general form of SFPCM for UT $k_q$ and UT $k'_{q'}$ as follows,
	\begin{align}
		\left [\mathbf{R}_{k'_{q'}}^{k_q}\right ]_{i,j}=
		\begin{cases}
			r_i\cdot\left[\mathbf{W}_{N_{\mathrm{c},\varphi _{k_{q}}-\varphi _{k'_{q'}}}}\right]_i,&i=j \\
			0,&\text{else}
		\end{cases},
	\end{align}
	where $r_i$ denotes the $i\text{th}$ diagonal element of matrix $\mathbf{R}^{q}_{q'}$, and it varies depending on the basic pilot matrices. Recalling (\ref{ADPCM}), the ADPCM of UT $k_q$ and UT $k'_{q'}$ can be summarized as
	\begin{equation}
		\begin{aligned}
			\left[\mathbf{Z}_{k'_{q'}}^{k_{q}}\right]_{i,j}
			=&\frac{1}{N_{\mathrm{c}}}\sum_{n=0}^{N_{\mathrm{c}-1}}r_n\cdot\left[\mathbf{W}_{N_{\mathrm{c},\varphi _{k_{q}}-\varphi _{k'_{q'}}}}\right]_{n}\\
			&\cdot \exp\left(-\frac{\bar{\imath}2\pi(i-j)n}{N_{\mathrm{c}}}\right).
		\end{aligned}
	\end{equation}
	Note that the matrix $\mathbf{Z}_{k'_{q'}}^{k_{q}}$ is a Toeplitz matrix, meaning that each element, except for those in the first row and column, is equal to its adjacent top left element. Given the characteristics of the Toeplitz matrix, we only need to calculate $[\mathbf{Z}_{k'_{q'}}^{k_{q}}]_{0,:}$ and $[\mathbf{Z}_{k'_{q'}}^{k_{q}}]_{:,0}$ in order to determine all elements of $\mathbf{Z}_{k'_{q'}}^{k_{q}}$.

	\begin{equation}\label{Z row}
		\begin{aligned}
			[\mathbf{Z}_{k'_{q'}}^{k_{q}}]_{0,j}
			=&\frac{1}{N_{\mathrm{c}}}\sum_{n=0}^{N_{\mathrm{c}-1}}\left[\operatorname{diag}\{\mathbf{R}_{k'_{q'}}^{k_q}\}\right]_{n}\cdot \exp\left(\frac{\bar{\imath}2\pi jn}{N_{\mathrm{c}}}\right)\\
			=&\left[\mathscr{F}^{-1}\left\{\operatorname{diag}\{\mathbf{R}_{k'_{q'}}^{k_q}\}\right\}\right]_{j}.
		\end{aligned}
	\end{equation}
	\begin{equation}\label{Z column}
		\begin{aligned}
			[\mathbf{Z}_{k'_{q'}}^{k_{q}}]_{i,0}
			=&\frac{1}{N_{\mathrm{c}}}\sum_{n=0}^{N_{\mathrm{c}-1}}r_n\cdot\left[\mathbf{W}_{N_{\mathrm{c},\varphi _{k_{q}}-\varphi _{k'_{q'}}}}\right]_{n}\cdot \exp\left(-\frac{\bar{\imath}2\pi in}{N_{\mathrm{c}}}\right)\\
			=&\frac{1}{N_{\mathrm{c}}}\left[\mathscr{F}\left\{\operatorname{diag}\{\mathbf{R}_{k'_{q'}}^{k_q}\}\right\}\right]_{i}.
		\end{aligned}
	\end{equation}
	From (\ref{Z column}) and (\ref{Z row}), it is evident that the elements in the first row and first column of $\mathbf{Z}_{k'_{q'}}^{k_{q}}$ are derived from the $N_{\mathrm{c}}$-point IDFT and DFT of the diagonal element $\mathbf{R}_{k'_{q'}}^{k_q}$ in the SFPCM matrix, respectively. This approach can be leveraged to efficiently compute all elements of ADPCM and aid in understanding the impact of ADPCM on pilot interference. 
	
	The following straightforward proposition presents an alternative perspective on the composition of ADPCM and offers an analysis of the underlying principle for alleviating pilot interference.
	
	\begin{theorem}\label{Theorem 1}
		The ADPCM of different basic pilot matrices displays characteristics of cyclic shifts with phase differences, as follows,	
		\begin{align}\label{eq:p1}
			\mathbf{Z}_{k'_{q'}}^{k_{q}}
			=&\frac{1}{N_{\mathrm{c}}}\mathbf{W}_{N_{\mathrm{c}}}^T\mathbf{D}_{\varphi _{k_q}-\varphi _{k'_{q'}}}\mathbf{R}^{q}_{q'}\mathbf{W}_{N_{\mathrm{c}}}^*\nonumber\\
			=&\frac{1}{N_{\mathrm{c}}}\mathbf{W}_{N_{\mathrm{c}}}^T\mathbf{D}_{\varphi _{k'_{q'}}-\varphi _{k_q}}\mathbf{W}_{N_{\mathrm{c}}}^*\mathbf{R}^{q}_{q'}\frac{\mathbf{W}_{N_{\mathrm{c}}}^T\mathbf{W}_{N_{\mathrm{c}}}^*}{\mathbf{W}_{N_{\mathrm{c}}}^T\mathbf{W}_{N_{\mathrm{c}}}^*}\nonumber\\
			=&\frac{1}{N_{\mathrm{c}}}\mathbf{W}_{N_{\mathrm{c}}}^T\mathbf{D}_{\varphi _{k'_{q'}}-\varphi _{k_q}}\mathbf{W}_{N_{\mathrm{c}}}^*\cdot\frac{1}{N_{\mathrm{c}}}\mathbf{W}_{N_{\mathrm{c}}}^T\mathbf{R}^{q}_{q'}\mathbf{W}_{N_{\mathrm{c}}}^*\nonumber\\
			=&\frac{1}{N_{\mathrm{c}}}\mathbf{W}_{N_{\mathrm{c}}}^T\mathbf{R}^{q}_{q'}\mathbf{W}_{N_{\mathrm{c}}}^*\mathbf{\Lambda }_{N_{\mathrm{c}}}^{\varphi _{k_q}-\varphi _{k'_{q'}}}\nonumber\\
			=&\frac{1}{N_{\mathrm{c}}}\mathbf{Z}^{q}_{q'}\mathbf{\Lambda }_{N_{\mathrm{c}}}^{\varphi _{k_q}-\varphi _{k'_{q'}}},
		\end{align}
		where
		\begin{equation}\label{basic ADPCM}
			\mathbf{Z}^{q}_{q'}=\mathbf{W}_{N_{\mathrm{c}}}^T\mathbf{R}^{q}_{q'}\mathbf{W}_{N_{\mathrm{c}}}^*,
		\end{equation}
		is the ADPCM of basic pilot matrices.
	\end{theorem}
	
	Proposition \ref{Theorem 1} provides a comprehensive expression for the pilot interference arising from MAPSP channel acquisition methods. The intensity of pilot interference primarily depends on the degree of channel overlap, the choice of the basic pilot matrices, and the phase shift differential. This means that the inter-group pilot interference matrix has a similar cyclic shift property as the intra-group pilot interference matrix. Recalling (\ref{diff group interference}), it can be restated as
	\begin{equation}\label{Diff groups after shift}
		\begin{aligned}
			\mathbf{H}_{k_q,\ell}^{\varphi_{k_{q}}-\varphi_{k'_{q'}}}
			&=N_{\mathrm{c}}\cdot\bar{\mathbf{H}}_{k_q,\ell}\mathbf{Z}_{k'_{q'}}^{k_q}\mathbf{I}_{N_{\mathrm{c}}\times N_{\mathrm{g}}}\\
			&=\bar{\mathbf{H}}_{k_q,\ell}\mathbf{Z}^{q}_{q'}\mathbf{\Lambda }_{N_{\mathrm{c}}}^{\varphi _{k_q}-\varphi _{k'_{q'}}}\mathbf{I}_{N_{\mathrm{c}}\times N_{\mathrm{g}}}.
		\end{aligned}
	\end{equation}
	
\begin{figure}[h!]
		\centering
		\includegraphics[width=1\linewidth]{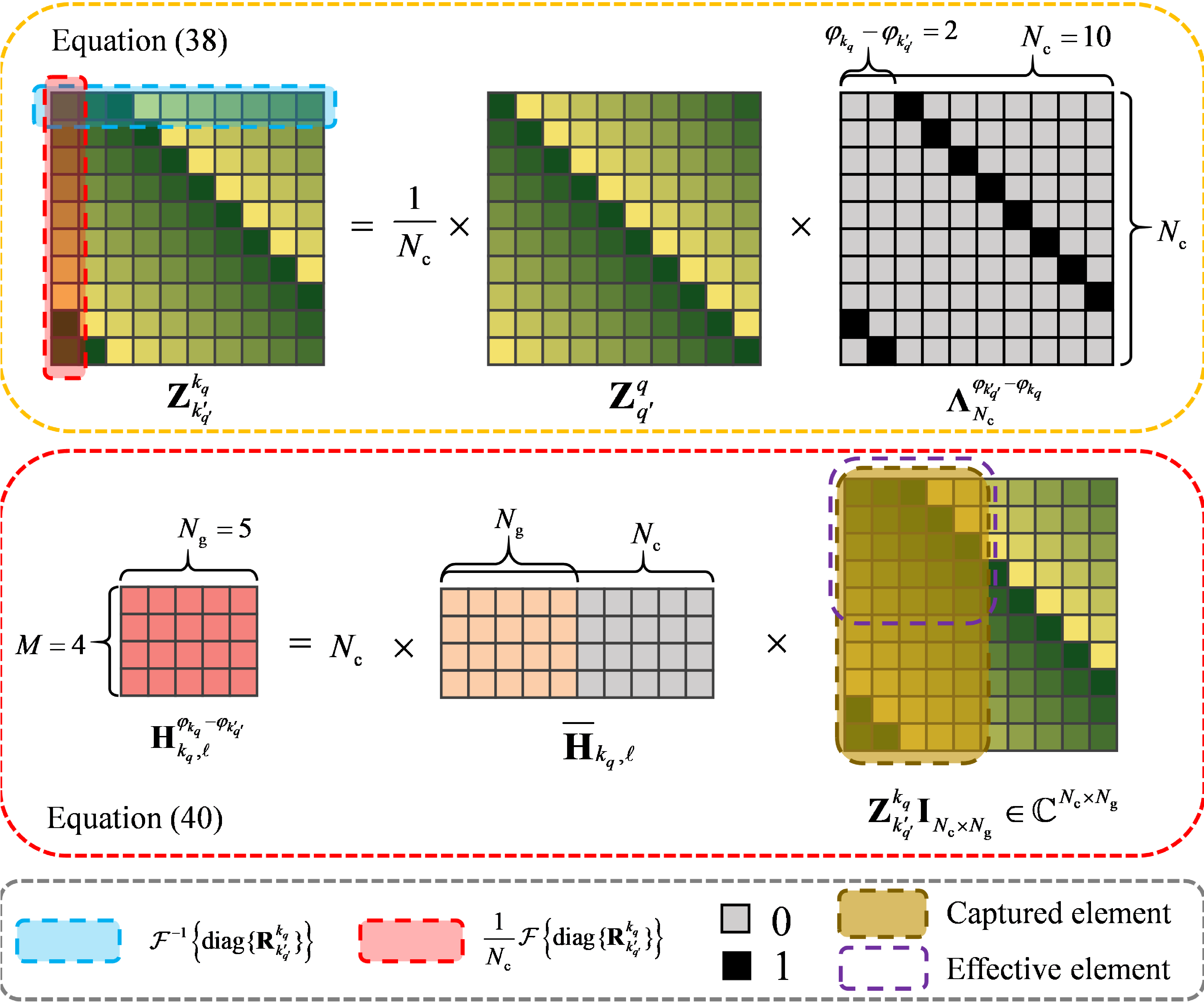}
		\captionsetup{font=footnotesize}
		\caption{Matrix structure diagram of pilot interference term.}\label{fig:proposition1}
	\end{figure}

	The interference between two groups arises from the ADPCM of their basic pilot matrix and $\mathbf{H}_{k_q,\ell}$. For the sake of description, we next analyze (\ref{Diff groups after shift}) in combination with \figref{fig:proposition1}, which contains the structure details of $\mathbf{H}_{k_q,\ell}^{\varphi_{k_{q}}-\varphi_{k'_{q'}}}$. The yellow and red dashed boxes in \figref{fig:proposition1} represent (\ref{eq:p1}) and (\ref{Diff groups after shift}), respectively. First, focus on the structure of $\mathbf{Z}^{k_q}_{k'_{q'}}$. The role of $\mathbf{\Lambda }_{N_{\mathrm{c}}}^{\varphi _{k_q}-\varphi _{k'_{q'}}}$ is to cyclically shift the matrix multiplied by it to the right, which means that $\mathbf{Z}^{k_q}_{k'_{q'}}$ can be viewed as $\mathbf{Z}^{q}_{q'}$ cyclically shifted by $\varphi _{k_q}-\varphi _{k'_{q'}}$ columns (see Proposition \ref{Theorem 1}). Since $\mathbf{\Lambda }_{N_{\mathrm{c}}}^{\varphi _{k_q}-\varphi _{k'_{q'}}}$ does not alter the structural properties of $\mathbf{Z}^{q}_{q'}$, $\mathbf{Z}^{k_q}_{k'_{q'}}$ is also a Toeplitz matrix. The blue and red dashed boxes in \figref{fig:proposition1} correspond to (\ref{Z row}) and (\ref{Z column}), respectively, indicating that in the Toeplitz matrix, $\mathscr{F}^{-1}\left\{\operatorname{diag}\{\mathbf{R}_{k'_{q'}}^{k_q}\}\right\}$ or $\frac{1}{N_{\mathrm{c}}}\mathscr{F}\{\operatorname{diag}\{\mathbf{R}_{k'_{q'}}^{k_q}\}\}$ encompass the complete information contained within $\mathbf{Z}^{k_q}_{k'_{q'}}$. Referring back to (\ref{Diff groups after shift}), $\mathbf{Z}_{k'_{q'}}^{k_q}\mathbf{I}_{N_{\mathrm{c}}\times N_{\mathrm{g}}}$ can be regarded as the result of truncating the first $N_{\mathrm{g}}$ columns of $\mathbf{Z}_{k'_{q'}}^{k_q}$. Consequently, we refer to the elements of the first $N_{\mathrm{g}}$ columns of $\mathbf{Z}_{k'_{q'}}^{k_q}$ as the captured elements (highlighted by the brown dashed box in \figref{fig:proposition1}). The matrix $\mathbf{H}_{k_q,\ell}^{\varphi_{k_{q}}-\varphi_{k'_{q'}}}$ is computed by multiplying $\bar{\mathbf{H}}_{k_q,\ell}$ with the captured elements. It is important to note that the last $(N_{\mathrm{c}}-N_{\mathrm{g}})$ columns of $\bar{\mathbf{H}}_{k_q,\ell}$ are all zeros, and thus the last $(N_{\mathrm{c}}-N_{\mathrm{g}})$ rows of $\mathbf{Z}_{k'_{q'}}^{k_q}\mathbf{I}_{N_{\mathrm{c}}\times N_{\mathrm{g}}}$, which are multiplied by these columns, do not affect the calculation result.
	
	In summary, only the first $N_{\mathrm{g}}$ rows of $\mathbf{Z}_{k'_{q'}}^{k_q}\mathbf{I}_{N_{\mathrm{c}}\times N_{\mathrm{g}}}$ are pertinent to the elements of $\mathbf{H}_{k_q,\ell}^{\varphi_{k_{q}}-\varphi_{k'_{q'}}}$. We refer these elements as the effective elements of $\mathbf{Z}^{k_q}_{k'_{q'}}$ (highlighted by the purple dashed box in \figref{fig:proposition1}). If all these $N_{\mathrm{g}}^2$ effective elements are zero, then the pilot interference will also be zero. It is important to note that each column of $\mathbf{Z}_{k'_{q'}}^{k_q}$ is merely a cyclic shift of the first column. Therefore, the values of the effective elements depend only on the first $N_{\mathrm{g}}$ and the last $N_{\mathrm{g}}$ elements of $[\mathbf{Z}_{k'_{q'}}^{k_{q}}]_{:,0}$. By referencing $\frac{1}{N_{\mathrm{c}}}\mathscr{F}\{\operatorname{diag}\{\mathbf{R}_{k'_{q'}}^{k_q}\}\}$, we can quickly calculate the pilot interference. Additionally, there is another scenario without pilot interference, which occurs when the non-zero pilot interference $\mathbf{H}_{k_q,\ell}^{\varphi_{k_{q}}-\varphi_{k'_{q'}}}$ does not overlap with the estimated $\mathbf{H}_{k_q,\ell}$. By scheduling the phase (changing $\mathbf{\Lambda }_{N_{\mathrm{c}}}^{\varphi _{k_q}-\varphi _{k'_{q'}}}$), we can adjust the positions of the elements in $[\mathbf{Z}_{k'_{q'}}^{k_{q}}]_{:,0}$ (thus, modifying $\mathbf{H}_{k_q,\ell}^{\varphi_{k_{q}}-\varphi_{k'_{q'}}}$) to mitigate pilot interference. Specifically, when the same basic pilot is deployed, $\mathbf{Z}^{q}_{q'}=N_{\mathrm{c}}\mathbf{I}_{N_{\mathrm{c}}}$, and (\ref{Diff groups after shift}) transforms into the intra-group interference (\ref{same group interference}).
	
	Based on the above analysis, we can summarize the following conclusions. 
	\begin{itemize}
		\item If $\mathbf{H}_{k_q,\ell}^{\varphi_{k_{q}}-\varphi_{k'_{q'}}}$ is a zero matrix, the fewer the non-zero elements in $\frac{1}{N_{\mathrm{c}}}\mathscr{F}\{\operatorname{diag}\{\mathbf{R}_{k'_{q'}}^{k_q}\}\}$, the more cases of phase shift exist where the first and last $N_{\mathrm{g}}$ elements of $[\mathbf{Z}_{k'_{q'}}^{k_{q}}]_{:,0}$ are zero.
		\item If $\mathbf{H}_{k_q,\ell}^{\varphi_{k_{q}}-\varphi_{k'_{q'}}}$ is a non-zero matrix, the fewer the non-zero elements among the effective elements, the lower the overlap between $\mathbf{H}_{k_q,\ell}^{\varphi_{k_{q}}-\varphi_{k'_{q'}}}$ and the estimated $\mathbf{H}_{k_q,\ell}$. 
	\end{itemize}
	Accordingly, we can intuitively derive the following design criteria for MAPSP.

	\begin{theorem}\label{Theorem 2}
		For all basic pilot matrices $\mathbf{S}_q$ that satisfy $\mathbf{S}_q\mathbf{S}_{q}^H=\mathbf{I}_{N_{\mathrm{c}}},\quad q\in \mathcal{Q}$, the optimal basic pilot matrices of MAPSP achieve the following minimization: for $\forall q,q'\in \mathcal{Q},\ q\neq q',\ \exists n\in \left\{0,1,\dots,N_{\mathrm{c}}-1\right\}$
		\begin{equation}
			\min\left\{\left|\frac{1}{N_{\mathrm{c}}}\mathscr{F}\left\{\operatorname{diag}\{\mathbf{R}^{q}_{q'}\}\right\}\right|-\boldsymbol{\delta}_n\right\},
		\end{equation}
		where
		\begin{equation}
			\left[\boldsymbol{\delta}_n\right]_i=\begin{cases}
				1,&i=n\\
				0,&\text{else}
			\end{cases}
		\end{equation}
		is the ideal solution of $|\frac{1}{N_{\mathrm{c}}}\mathscr{F}\{\operatorname{diag}\{\mathbf{R}^{q}_{q'}\}\}|$.
	\end{theorem}
	
\begin{figure}[h!]
		\centering
		\subfigure[$\mathbf{s}_q=\mathbf{s}_{q'}=\mathbf{W}_{2048,0}$]{
			\centering
			\includegraphics[width=1\linewidth]{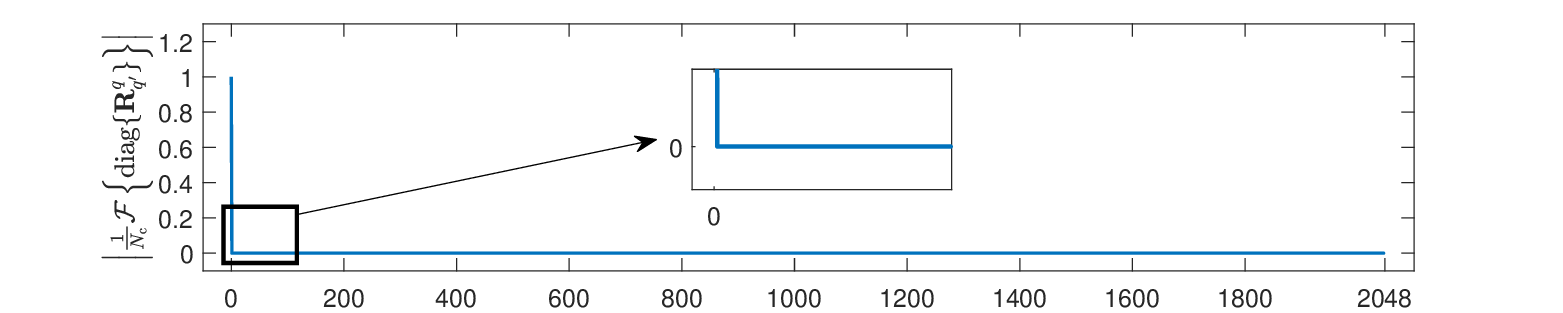}
			\captionsetup{font=footnotesize}
			\label{DFT_R_DFT_column}}
			
		\centering
		\subfigure[$\mathbf{s}_q=\boldsymbol{\chi}_{2048,200}^{1},\ \mathbf{s}_{q'}=\text{\large[}\text{[}\boldsymbol{\chi}_{2048,0}^{1}\text{]}_{0:1023},\text{[}\boldsymbol{\chi}_{2048,0}^{1}\text{]}_{0:1023}\text{\large]}$]{
			\centering
			\includegraphics[width=1\linewidth]{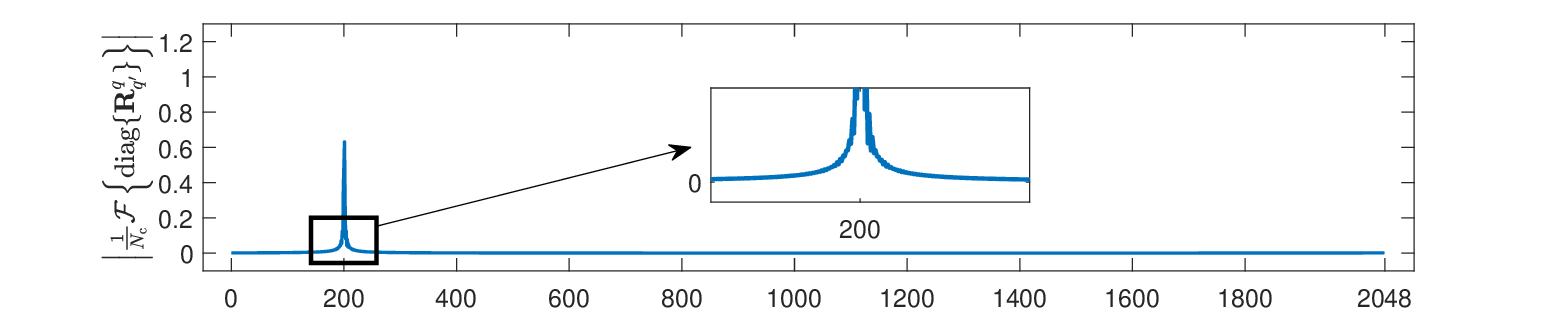}
			\captionsetup{font=footnotesize}
			\label{DFT_R_ZC_diff200}}
		
		\centering
		\subfigure[$\mathbf{s}_q=\boldsymbol{\chi}_{2048,0}^{1},\ \mathbf{s}_{q'}=\boldsymbol{\chi}_{2048,200}^{1}$]{
			\centering
			\includegraphics[width=1\linewidth]{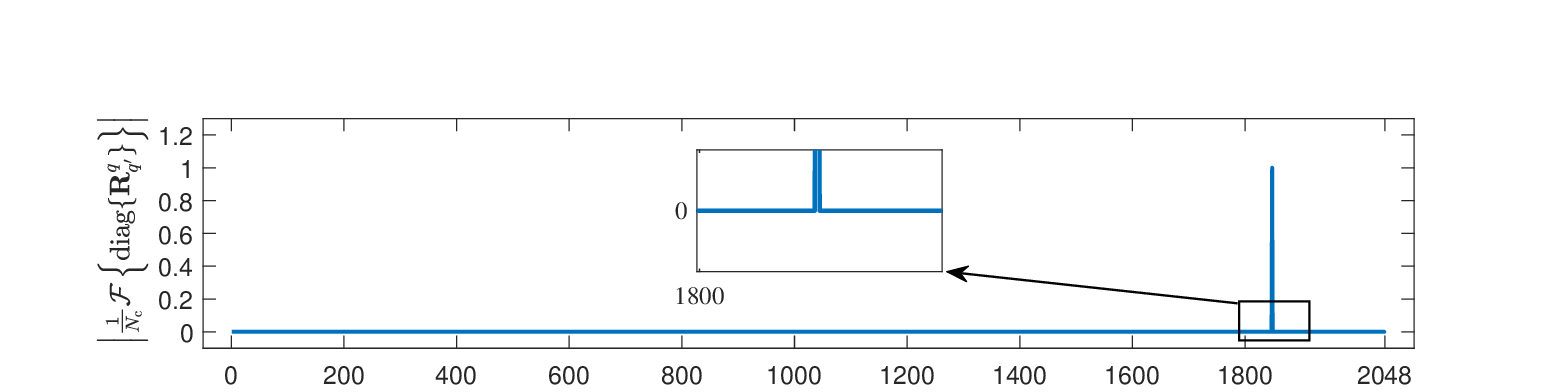}
			\captionsetup{font=footnotesize}
			\label{ZC_R_ZC}}
		
		\centering
		\subfigure[$\mathbf{s}_q=\boldsymbol{\chi}_{2048,0}^{1},\ \mathbf{s}_{q'}=\boldsymbol{\chi}_{2048,0}^{11}$]{
			\centering
			\includegraphics[width=1\linewidth]{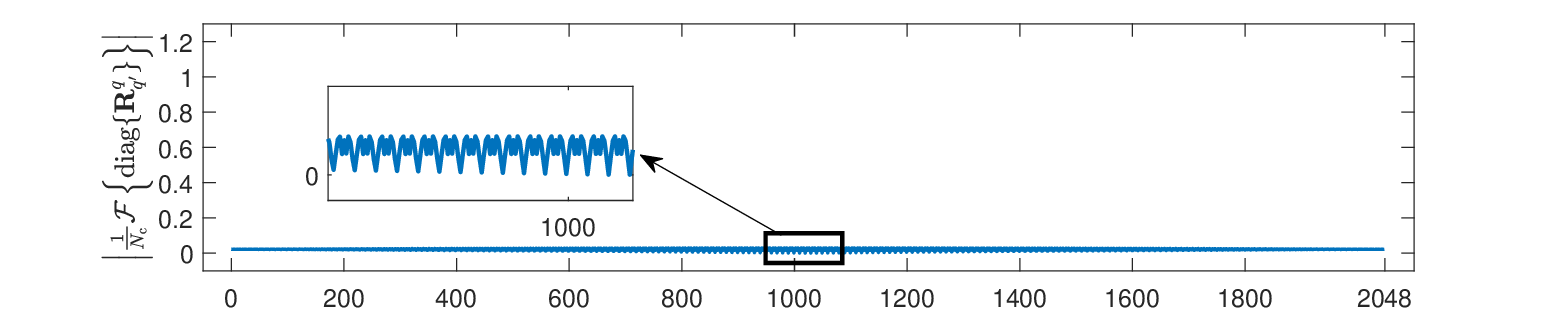}
			\captionsetup{font=footnotesize}
			\label{DFT_R_ZC_root1and11}}
			
		\captionsetup{font=footnotesize}
		\caption{The absolute values of first column elements of ADPCM generated by dual basic pilot matrices}
		\label{fig:dual basic mat}
	\end{figure}

	Proposition \ref{Theorem 2} states that the more similar the $|\frac{1}{N_{\mathrm{c}}}\mathscr{F}\{\operatorname{diag}\{\mathbf{R}^{q}_{q'}\}\}|$ and $\boldsymbol{\delta}_n$ of basic pilot matrices are, the better suited they are to generating MAPSP. The explanation of Proposition \ref{Theorem 2} is straightforward. All the other elements in $\boldsymbol{\delta}_n$ are zeros, except only one element of $1$, making it easy to shift $\boldsymbol{\delta}_n$ into $\boldsymbol{\delta}_{n'},\ n'\in \{N_{\mathrm{g}},N_{\mathrm{g}}+1,\dots,N_{\mathrm{c}}-N_{\mathrm{g}}-1\}$ with the help of phase shifting. The first and last $N_{\mathrm{g}}$ elements of $\boldsymbol{\delta}_{n'}$ are zeros, satisfying the condition of no pilot interference. Alternatively, since sparse $\boldsymbol{\delta}_n$ has only a single non-zero element, it helps maintain the original sparsity of the channel and facilitates simple pilot scheduling. When employing one basic pilot matrix to generate APSP, that is $|\frac{1}{N_{\mathrm{c}}}\mathscr{F}\{\operatorname{diag}\{\mathbf{R}^{q}_{q'}\}\}|=\boldsymbol{\delta}_0$, the sparsity of the channel is fully exploited. \figref{fig:dual basic mat} displays $|\frac{1}{N_{\mathrm{c}}}\mathscr{F}\{\operatorname{diag}\{\mathbf{R}^{q}_{q'}\}\}|$ for several $\mathbf{s}_q$ as a detailed explanation of Proposition \ref{Theorem 2}.
	
	Denote the ZC sequence as follows
	\begin{equation}\label{ZC}
		\left[\boldsymbol{\chi}_{N,\phi}^\mathfrak{r}\right]_n=\exp\left(-\bar{\imath}\pi\frac{\mathfrak{r}\cdot \mathscr{N}(\mathscr{N}+1)}{N}\right),
	\end{equation}
	\begin{equation}\label{ZC mod}
		\mathscr{N}=\left\langle n-\phi\right\rangle_{N-1},
	\end{equation}
	where $\mathfrak{r}$ denotes the root index of the ZC sequence, $N$ denotes the length of the sequence, and $\phi\in[0,1,\cdots,N-2]$ denotes the number of cyclic shifts of the ZC sequence itself. \figref{DFT_R_DFT_column}, \figref{DFT_R_ZC_diff200}, and \figref{DFT_R_ZC_root1and11} demonstrate three sparsity conditions: ideal sparsity, partial sparsity, and no sparsity, respectively. The total pilot interference is measured by $\sum_i\left[\left|\frac{1}{N_{\mathrm{c}}}\mathscr{F}\left\{\operatorname{diag}\{\mathbf{R}^{q}_{q'}\}\right\}\right|\right]_i$.
	
	In \figref{DFT_R_DFT_column}, where the same basic pilot matrix exhibits autocorrelation, $\left|\frac{1}{N_{\mathrm{c}}}\mathscr{F}\left\{\operatorname{diag}\{\mathbf{R}^{q}_{q'}\}\right\}\right|$, contains only one non-zero element, and $\sum_i\left[\left|\frac{1}{N_{\mathrm{c}}}\mathscr{F}\left\{\operatorname{diag}\{\mathbf{R}^{q}_{q'}\}\right\}\right|\right]_i=1$. \figref{DFT_R_ZC_diff200} presents a partially sparse scenario with multiple non-zero elements. Despite lower peak values than that in \figref{DFT_R_DFT_column}, the sum exceeds 1, indicating that the overall pilot interference is larger. \figref{DFT_R_ZC_root1and11} shows non-sparsity with all elements being non-zero, its total interference reached $\sum_i\left[\left|\frac{1}{N_{\mathrm{c}}}\mathscr{F}\left\{\operatorname{diag}\{\mathbf{R}^{q}_{q'}\}\right\}\right|\right]_i=43.69$. It is worth noting that, even though the peak height of the non-zero segments in \figref{DFT_R_ZC_root1and11} is not substantial, each UT exerts interference on the others, leading to a detrimental impact on the whole. The case in \figref{ZC_R_ZC}, which exhibits ideal cross-correlation properties, will be discussed in detail in Section \ref{sec_implementation}. Analysis of the curves in \figref{fig:dual basic mat} reveals an observation regarding sparsity characteristics: When $\sum_i\left[\left|\frac{1}{N_{\mathrm{c}}}\mathscr{F}\left\{\operatorname{diag}\{\mathbf{R}^{q}_{q'}\}\right\}\right|\right]_i$ is larger, the number and width of the non-zero segments of $|\frac{1}{N_{\mathrm{c}}}\mathscr{F}\{\operatorname{diag}\{\mathbf{R}^{q}_{q'}\}\}|$ are also larger, resulting in stronger pilot interference.
	
	\section{MAPSP Implementation Scheme}\label{sec_implementation}
	
	In this section, we propose an implementation of MAPSP based on ZC sequences. Leveraging the cross-correlation properties of ZC sequences, ADPCM offers a priori PAI. Correspondingly, we propose a received signal pre-processing method and phase scheduling algorithm to alleviate inter-group and intra-group pilot interference, respectively.
	
	\subsection{Received Signal Pre-processing Method}
	
	When multiple basic pilot matrices are utilized, it is equivalent to having advanced knowledge of the ADPCM information. By effectively combining channel characteristics and utilizing PAI, the performance of channel estimation can be enhanced even further. Given the known characteristics of $\mathbf{Z}^{q}_{q'}$, it can potentially mitigate or minimize pilot interference from the other groups. When combined with pilot scheduling, the prior information on the channel and pilot can be simultaneously utilized to aid channel estimation. 
	
	According to the optimal conditions for designing MAPSP in Proposition \ref{Theorem 2}, we use the ZC and its cyclic shift sequences as the basic pilot sequences. When $\mathbf{s}_{q'}=\boldsymbol{\chi}_{N_\mathrm{c},0}^\mathfrak{r}$, $\mathbf{s}_q=\boldsymbol{\chi}_{N_\mathrm{c},\phi}^\mathfrak{r}$, the angle-delay domain cross-correlation matrix of $\mathbf{Z}^{q}_{q'}$ is
	\begin{equation}\label{basic cro corr}
		\left[\mathbf{Z}^{q}_{q'}\right]_{i,j}\!\!\!=\!\!\begin{cases}
			\!\! N_\mathrm{c}\exp\!\left(-\bar{\imath}\pi\mathfrak{r} \frac{\phi(\phi-1)}{N_{\mathrm{c}}}\right)\!\!\!\!\!\!\!&,\left\langle i-j, N_\mathrm{c}\right\rangle\!\!=\!\!\left\langle \mathfrak{r}\phi\right\rangle_{N_\mathrm{c}},\\
			\!\!0&,\mathrm{else}.
		\end{cases}
	\end{equation}
	The proof of (\ref{basic cro corr}) is presented in Appendix \ref{appendix a}.
	
	When a ZC sequence and its cyclic shift sequences are employed as the basic pilot matrices, the condition $|\frac{1}{N_{\mathrm{c}}}\mathscr{F}\{\operatorname{diag}\{\mathbf{R}^{q}_{q'}\}\}|=\boldsymbol{\delta}_\phi$ can be met. It is worth noting that unlike the basic pilot sequence with autocorrelation $\mathbf{S}_q\mathbf{S}_{q}^H=\mathbf{I}_{N_{\mathrm{c}}}$, the $\mathbf{Z}^{q}_{q'}$ generated by the cross-correlation of different sequences here inherently contains argument information $\Theta ^{q}_{q'}$,
	\begin{equation}
		\Theta^{q}_{q'}=-\mathfrak{r} \frac{\phi(\phi-1)}{N_{\mathrm{c}}}\pi.
	\end{equation}
	We denote $\Theta^{q'}_{q}$ as the prior PAI of the $q'\text{th}$ group UT for $q\text{th}$ group, with the relationship $\Theta^{q'}_{q}=-\Theta^{q}_{q'}$. Substituting (\ref{basic cro corr}) into (\ref{Diff groups after shift}), pilot interference terms are reformulated,
	\begin{equation}\label{eq:inter group interference}
		\mathbf{H}_{k_q,\ell}^{\varphi_{k_{q}}-\varphi_{k'_{q'}}}=N_\mathrm{c}\exp\left(\bar{\imath}\Theta^{q'}_{q}\right)\cdot\bar{\mathbf{H}}_{k_q,\ell}\mathbf{\Lambda }_{N_{\mathrm{c}}}^{\varphi_{k_{q}}-\varphi_{k'_{q'}}+\phi}\mathbf{I}_{N_{\mathrm{c}}\times N_{\mathrm{g}}}.
	\end{equation}
	We can observe that the interference between different groups is characterized by a coefficient $\exp(\bar{\imath}\Theta^{q'}_{q})$, as opposed to the interference within the same group. Considering (\ref{h model}), the argument of $[\mathbf{H}_{k_q,\ell}]_{i,j}$ is a random variable $\bar{\theta}_{i,j}^{k_q}$. $\bar{\mathbf{H}}_{k_q,\ell}$ is the extended version of $\mathbf{H}_{k_q,\ell}$ after zero-padding, so its arguments also follow the same Gaussian distribution. Refer to (\ref{eq:inter group interference}), the presence of PAI $\Theta^{q}_{q'}$ has changed the arguments distribution mean value of $\bar{\mathbf{H}}_{k_q,\ell}$, so the arguments of $\exp\left(\bar{\imath}\Theta^{q'}_{q}\right)\bar{\mathbf{H}}_{k_q,\ell}$ (and 
	$\mathbf{H}_{k_q,\ell}^{\varphi_{k_{q}}-\varphi_{k'_{q'}}}$) follow Gaussian distribution $\mathcal{N}(\bar{\mu}_{i,j}^{k_q}+\Theta^{q}_{q'},\bar{\sigma}^2)$. In other words, $\Theta^{q'}_{q}$ staggers the arguments of the two groups of UT channels, and the concentrated range has only a few overlaps. Consequently, the intra-group and inter-group interference exhibit different statistical characteristics. Leveraging this property of MAPSP, we propose a pre-processing method to alleviate inter-group pilot interference. 
	
	First, we define the return-to-zero argument matrix of the $k_q\text{th}$ UT as $[\mathbf{\Theta}_{0}^{k_q}]_{i,j}=\exp(-\bar{\imath}\bar{\mu}^{k_q}_{i,j})$. Next, we compute the equivalent PAI matrix $\mathbf{\Theta}_{\Sigma}^{q'}$ between the $q'\text{th}$ group UT channel and the other group channels.
	\begin{align}
		\mathbf{\Theta}_{\Sigma}^{q'}\!=\!\arg\big\{\!\!\sum_{q\in\mathcal{Q}\backslash q'}\!\!\!\!\exp(\bar{\imath}\Theta^{q'}_{q})\!\!\sum_{k_q=0}^{K_q-1}\!\!\mathbf{P}_{k_q}^{\varphi_{k_{q}}-\varphi_{k'_{q'}}}\odot\mathbf{\Theta}_{0}^{k_q}\big\},
	\end{align}
	where $\arg\{\cdot\}$ represents the calculation argument operation. Then, the received signal after pre-processing is as follows, 
	\begin{equation}
		\mathbf{\Theta}_{0}^{k'_{q'}}\!\!\!\odot\!\!\breve{\mathbf{Y}}_{k'_{q'},\ell}=\Re\left\{\mathbf{\Theta}_{0}^{k'_{q'}}\!\!\!\odot\!\!\mathbf{Y}_{k'_{q'},\ell}\right\} -\frac{\Im\left\{\mathbf{\Theta}_{0}^{k'_{q'}}\!\!\!\odot\!\!\mathbf{Y}_{k'_{q'},\ell}\right\}}{\tan (\mathbf{\Theta}_{\Sigma}^{q'}) }.
	\end{equation}
	From a statistical perspective, the pre-processed signal $\breve{\mathbf{Y}}_{k'_{q'},\ell}$ effectively alleviates interference from the groups,
	\begin{equation}\label{elimination}
		\mathsf{E}\left\{\mathbf{\Theta}_{0}^{k'_{q'}}\!\!\!\odot\!\!\left[\breve{\mathbf{Y}}_{k'_{q'},\ell}\right]_{i,j}\right\}=\left[\sum_{k_{q'}\in  \mathcal{K}_{q'}}\sqrt{\mathbf{P}_{k_{q'},\ell}^{\varphi_{k_{q'}}-\varphi_{k'_{q'}}}}\!\right]_{i,j}\!\!\!\!\!\exp(-\frac{\bar{\sigma}^2}{2}).
	\end{equation}
	The proof of (\ref{elimination}) is presented in Appendix \ref{appendix b}.

	Finally, we take the pre-processed $\mathbf{\Theta}_{0}^{k'_{q'}}\!\!\!\odot\!\breve{\mathbf{Y}}_{k'_{q'},\ell}/\exp(-\frac{\bar{\sigma}^2}{2})$ as the received signal of BS and perform MMSE channel estimation. After pre-processing, the interference between groups is effectively alleviated. Under the premise of accommodating more UTs, the proposed MAPSP method has estimation errors similar to those of APSP. However, it is important to note that increasing the number $Q$ of groups also increases the difficulty of reducing inter-group interference, as confirmed in subsequent comprehensive numerical simulations. These errors are attributed to the incomplete elimination of inter-group interference, which comes from the distribution assumption of channel argument. When the channel parameter $\bar{\sigma}^2$ decreases, the error caused by pre-processing is smaller.
	
	\subsection{Phase Scheduling Algorithm}
	
	After conducting the necessary mathematical verification, we have obtained the MMSE channel estimation of MAPSP in massive MIMO-OFDM, along with the identification of the causes of intra-group and inter-group pilot interference terms. To address intra-group interference, leveraging the channel sparsity of the angle-delay domain can effectively reduce or eliminate channel overlap, thereby minimizing pilot interference. It's worth noting that the intensity of the interference is influenced by the power distribution of each UT channel, making pilot scheduling a crucial and advantageous practice \cite{you2015channel}. As for inter-group interference, we have developed MAPSP with PAI based on Proposition \ref{Theorem 2}. When the argument distribution follows Gaussian distribution, $\Theta^{q'}_{q}$ can adjust the argument distribution range of two groups of UT channels. This adjustment allows the BS to reduce inter-group pilot interference through pre-processing.
	
	Pilot scheduling employs the intra-group MMSE criterion. As the phase shift of one channel impacts the other UTs, the optimal scheme is derived by iterating through all cases in intra-group scheduling. Therefore, we propose a MAPSP scheduling algorithm to simplify the calculation by setting intra-group threshold $\Upsilon_\mathrm{tra}$, even if it means some scheduling effect sacrifices. 
	
	We utilize the normalized Hadamard product of $\mathbf{P}_{k_q}$ to denote the degree of overlaps between channels. The channel superposition matrix $\bar{\mathbf{P}}_\Sigma^q$ is used to describe the overlaps of the scheduled channels in the $q\text{th}$ group. The grouping and the phase shift factor are assigned to each UT one by one. When no phase meets the threshold $\Upsilon_\mathrm{tra}$, the factor with the most minor overlap is selected. In our algorithm, each UT is scheduled in all groups successively and finally assigned to the group with the minimum overlap, which takes further advantage of the inherent channel sparsity.
	
	The intra-group scheduling iterations follow a uniform distribution over $[1,N_\mathrm{c}-1]$, resulting in an average computational complexity of $\mathcal{O}(KQ(N_\mathrm{c}+1)/2)$. In contrast to the exhaustive greedy search approach with complexity $\mathcal{O}(KQN_\mathrm{c})$, our proposed method employs an adaptive threshold mechanism to achieve a balanced trade-off between scheduling performance and computational overhead. Specifically, when $\Upsilon_\mathrm{tra}=0$, the algorithm degenerates to the conventional greedy method, attaining optimal scheduling performance at the expense of higher computational cost. By contrast, increasing the threshold $\Upsilon_\mathrm{tra}$ leads to progressive reduction in computational complexity while accepting corresponding degradation in scheduling performance. This design provides an adjustable operational flexibility to meet diverse practical requirements. The proposed scheduling algorithm is detailed in Algorithm \ref{alg:MAPSP}.
	
	\begin{figure}[h!]
		\begin{algorithm}[H]
			\caption{MAPSP Scheduling Algorithm}
			\label{alg:MAPSP}
			\begin{algorithmic}[1]
				\Require UT sets $\mathcal{K}_q$ and $q\in \mathcal{Q}$; the channel power matrix  $\mathbf{P}_{k_q}$, $k_q \in \mathcal{K}_q$; intra-group scheduling thresholds $\Upsilon_\mathrm{tra}$
				\Ensure Pilot phase shift pattern $\varphi_{k_q}$; updated UT sets $\mathcal{K}_q$
				\State {\textbf{Step 1 : Initialization}}
				\State {A UT set to be scheduled $\mathcal{K}^{\mathrm{temp}}=\mathcal{K}_0\backslash\{0\}\cup \mathcal{K}_1\backslash\{0\}\cup\cdots\cup\mathcal{K}_{Q-1}\backslash\{0\}$, and empty UT sets $\mathcal{K}_q=\{0\}$; $\varphi_{0_q}=0$; the channel superposition matrix $\bar{\mathbf{P}}_\Sigma^{q}=\bar{\mathbf{P}}_{0_q}$}
				\State {\textbf{Step 2 : MAPSP scheduling}}
				\State{Randomly select a UT $k$ from the set $\mathcal{K}^\mathrm{temp}$}
				\For{$q\in\mathcal{Q}$}
				\State{Schedule in the $q\text{th}$ group and set $\Gamma_\mathrm{min}^q=\inf$}
				\For{$\varphi=0,1,\cdots,N_{\mathrm{c}}-1$}
				\If{the number of UT in $\mathcal{K}_q$ is equal to $K_q$}
				\State{\textbf{break}}
				\EndIf
				\State{$\Gamma_\mathrm{temp}^q=\sum_{i,j}\left[\bar{\mathbf{P}}_{k}\mathbf{\Lambda }_{N_{\mathrm{c}}}^{\varphi}\odot\bar{\mathbf{P}}_\Sigma^{q}\right]_{i,j}$}
				\State{$\Gamma_\Sigma=\Upsilon_\mathrm{tra}\sqrt{\sum_{i,j}\left[\bar{\mathbf{P}}_{k}\mathbf{\Lambda }_{N_{\mathrm{c}}}^{\varphi}\right]_{i,j}\sum_{i,j}\left[\bar{\mathbf{P}}_\Sigma^{q}\right]_{i,j}}$}
				\If{$\Gamma_\mathrm{temp}^q\le\Gamma_\Sigma$}$\varphi^q=\varphi$, $\Gamma_\mathrm{min}^q=\Gamma_\mathrm{temp}^q$,  \textbf{break}
				\ElsIf{$\Gamma_\mathrm{temp}^q<\Gamma_\mathrm{min}^q$}$\varphi^{q}=\varphi$,$\Gamma_\mathrm{min}^q=\Gamma_\mathrm{temp}^q$
				\EndIf
				
				\EndFor
				\EndFor
				\State {\textbf{Step 3 : Update UT sets and phase shift}}
				\State{Compare to find the smallest $\Gamma_\mathrm{min}^q$, and assign UT $k$ to the $q\text{th}$ set. $\mathcal{K}^\mathrm{temp}=\mathcal{K}^\mathrm{temp}\backslash k$}
				\State{$\mathcal{K}_q=\{k\}\cup\mathcal{K}_q$, $\varphi_{k_q}=\varphi^q$, $\bar{\mathbf{P}}_\Sigma^{q}=\bar{\mathbf{P}}_\Sigma^{q}+\bar{\mathbf{P}}_{k}\mathbf{\Lambda }_{N_{\mathrm{c}}}^{\varphi_{k_q}}$}
				\State{Repeat \textbf{Step 2} until $\mathcal{K}^\mathrm{temp}=0$}
			\end{algorithmic}
		\end{algorithm}
	\end{figure}
	
	\section{Numerical Results}\label{sec_simulation}
	
	In this section, we evaluate the performance of the proposed MAPSP channel acquisition method in massive MIMO-OFDM through numerical simulations. We utilize the OFDM parameter configuration from 3GPP LTE \cite{3gpp36.211} and summarize the detailed OFDM system parameters in \tabref{tb:OFDM parameters}. For generating channel information, we employ the professional radio channel generation software QuaDRiGa \cite{quadriga} to produce the UT's channel information. The simulation experiment considers three massive MIMO-OFDM scenarios specified in 3GPP 38.901 \cite{3gpp38.901}, namely Urban Macro (UMa), Urban Micro (Umi)-street canyon, and Indoor-office, with the corresponding parameter configurations listed in \tabref{tb:scenarios parameters}. For simplicity, the following UMi refers to the Umi-street canyon.
	 
	\newcolumntype{L}{>{\hspace*{-\tabcolsep}}l}
	\newcolumntype{R}{c<{\hspace*{-\tabcolsep}}}
	\definecolor{lightblue}{rgb}{0.93,0.95,1.0}
	\definecolor{lightgreen}{rgb}{0.95,1.0,0.93}
	\begin{table}[htbp]
		\captionsetup{font=footnotesize}
		\caption{OFDM System Parameters}\label{tb:OFDM parameters}
		\centering
		\ra{1.5}
		\scriptsize
		\begin{tabular}{LR}
			\toprule
			Parameter &  Value \\
			\rowcolor{lightblue}
			\midrule
			Bandwidth & 20 MHz\\
			Subcarrier spacing &15 KHz \\
			\rowcolor{lightblue}
			Sample interval $T_s$
			&32.6 ns \\
			Subcarrier number $N_{\mathrm{c}}$&2048 \\
			\rowcolor{lightblue}
			CP length $N_{\mathrm{g}}$ & 144 \\
			Symbol duration $T_{\mathrm{sym}}$ & 71.4 $\mu $s \\
			\bottomrule
		\end{tabular}
	\end{table}
	
	\newcolumntype{L}{>{\hspace*{-\tabcolsep}}l}
	\newcolumntype{R}{c<{\hspace*{-\tabcolsep}}}
	\definecolor{lightblue}{rgb}{0.93,0.95,1.0}
	\definecolor{lightgreen}{rgb}{0.95,1.0,0.93}
	\begin{table}[htbp]
		\captionsetup{font=footnotesize}
		\caption{Generate Channel Parameters in Typical Scenarios}\label{tb:scenarios parameters}
		\centering
		\ra{1.5}
		\scriptsize
		\begin{tabular}{LccR}
			\toprule
			Scenario &  UMa &UMi &Indoor-office \\
			\rowcolor{lightblue}
			\midrule
			Center frequency & 6 GHz &6 GHz&6 GHz\\
			Antenna downtilt &102$^{\circ}$ &102$^{\circ}$ &110$^{\circ}$\\
			\rowcolor{lightblue}
			BS height &25 m &10 m & 3 m\\
			BS transmit power &49 dB &44 dB &24 dB\\
			\rowcolor{lightblue}
			Spatial correlation distance &50 m &15 m &10 m \\
			Translational speed &80 Km/h &40 Km/h &5 Km/h \\
			\rowcolor{lightblue}
			Doppler $\nu T_{\mathrm{sym}}$ &31.4$\times$10$^{-3}$ &16.2$\times$10$^{-3}$ &2.52$\times$10$^{-3}$ \\
			Path standard deviation $\bar{\sigma}$ &-10 dB&-10 dB&-10 dB\\
			\bottomrule
		\end{tabular}
	\end{table}
	
	We focus on a single hexagonal cell with all UTs served within the cell. There is a single BS positioned at the geometric center of the hexagonal cell, equipped with a ULA consisting of 128 antennas. Assume that the UTs are randomly distributed within the $60^{\circ}$ range in the middle of the three sectors of the BS, and each UT's AOA is also automatically generated by QuaDRiGa. All UTs are situated at a height of 1.5m. In the context of the delay domain, the number of channel taps generated by QuaDRiGa is variable but exceeds 17 for all channels. To streamline the simulation, we make sure that the guard interval length $T_{\mathrm{g}}=N_{\mathrm{g}}T_{\mathrm{s}}$ is greater than the maximum channel delay of all UTs. It is assumed that all UT transmissions are synchronous. $\bar{\sigma}$ is determined by the similarity between the phases of the NLoS and LoS components, we assume that the phase similarity is the same for all three scenes. The Doppler frequency parameter $\nu$ of the channel is selected as the frequency with the highest power from the Doppler power spectrum of a UT that is traveling at a constant speed based on Monte Carlo simulations under their respective scenarios. For simulation convenience, it is assumed that all UTs share the same Doppler frequency within the same scenario. In the numerical simulation, the large-scale fading of channels is not considered, and all $K$ UT channels are normalized to $\sum_{i,j}\left[\boldsymbol{P}_{k_q}\right]_{i,j}=MN_{\mathrm{c}}$ for simplicity. 
	
	Using the parameter configurations from the three scenarios mentioned above and considering the APSP and CS methods as the performance benchmark, we compare the MMSE error and SE metrics of three channel information acquisition methods for $K=42$, $K=84$ and $K=126$. When using the CS channel estimation algorithm, the UT transmits PSOPs. To achieve accurate channel estimation for all the 42 UTs, the CS algorithm requires pilot transmission over three OFDM symbols. Thus, CS performance for larger UT counts is not simulated. The remaining simulations are all conducted on a single OFDM symbol. Divide the simulation into three sets based on the number of UTs. In each set, we successively use two methods to estimate the same $K$ UT channels. The threshold in each simulation of both methods is $\Upsilon_\mathrm{tra}=10^{-7}$. In the simulation results, 2-APSP denotes the MAPSP scheme for two UT groups, 3-APSP represents the MAPSP for three UT groups, and PSOP indicates the CS method.
	
	\begin{figure*}[!t]
		\centering
		\subfigure[UMa]{
			\centering
			\includegraphics[width=0.31\linewidth]{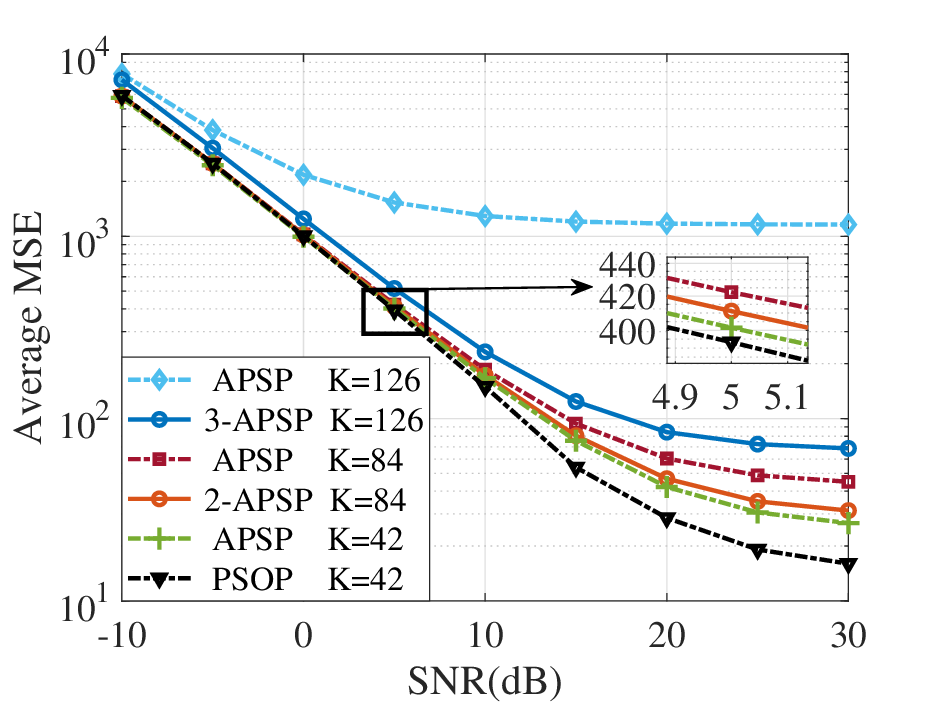}
			\captionsetup{font=footnotesize}
			\label{UMa_MMSE}}
		\subfigure[UMi]{
			\centering
			\includegraphics[width=0.31\linewidth]{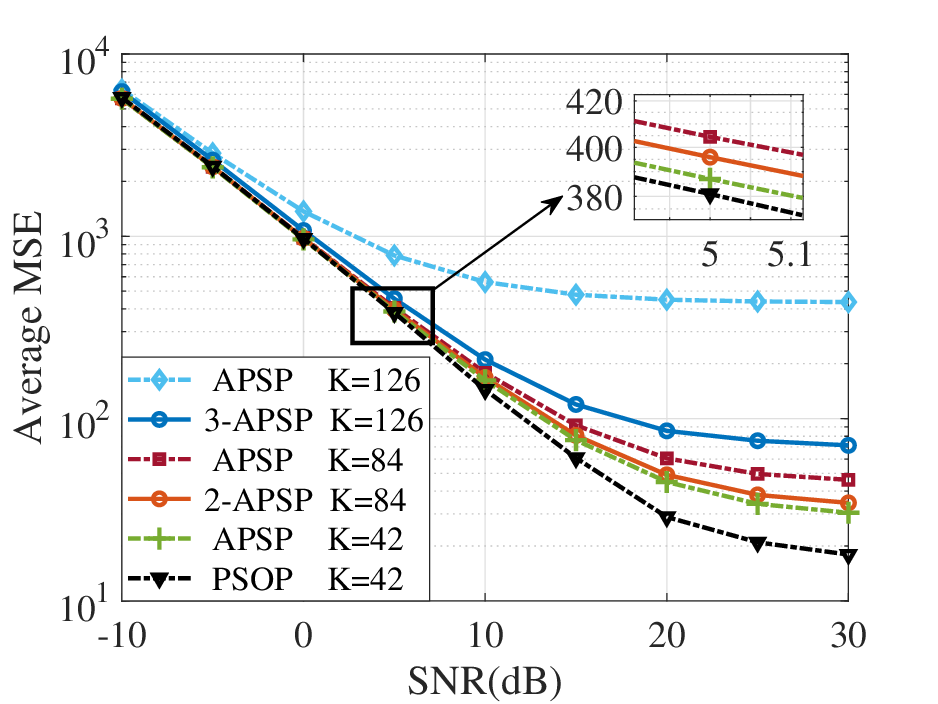}
			\captionsetup{font=footnotesize}
			\label{UMi_MMSE}}
		\subfigure[Indoor-office]{
			\centering
			\includegraphics[width=0.31\linewidth]{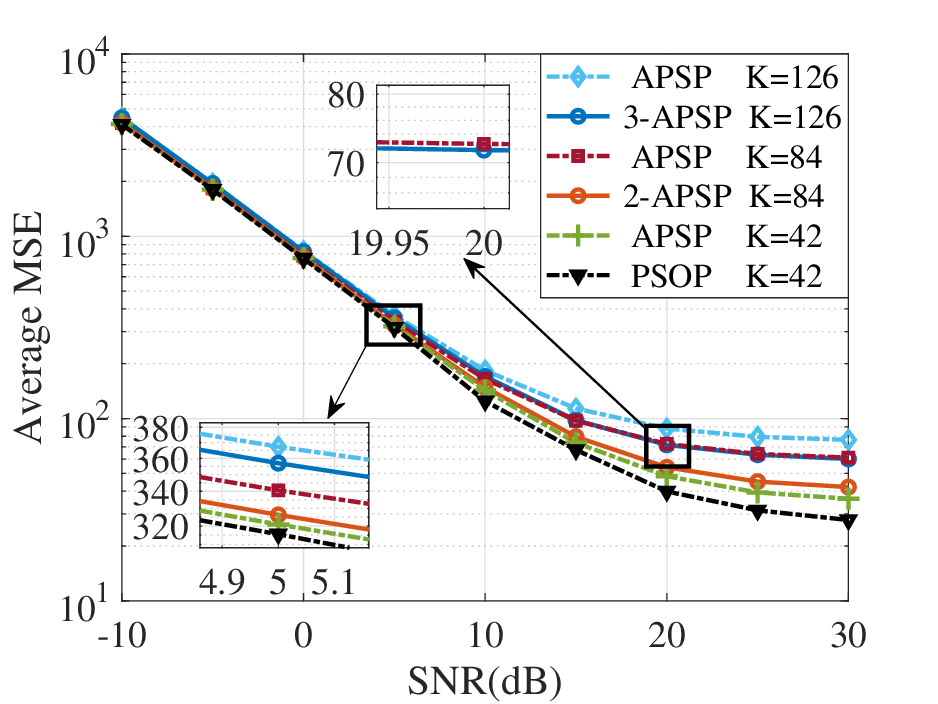}
			\captionsetup{font=footnotesize}
			\label{Id_MMSE}}
		
		\captionsetup{font=footnotesize}
		\caption{Comparison of channel estimation MMSE by APSP and MAPSP under different UT numbers}
		\label{fig:MMSE}
	\end{figure*}
	
	\begin{figure*}[!t]
		\centering
		\subfigure[UMa]{
			\centering
			\includegraphics[width=0.31\linewidth]{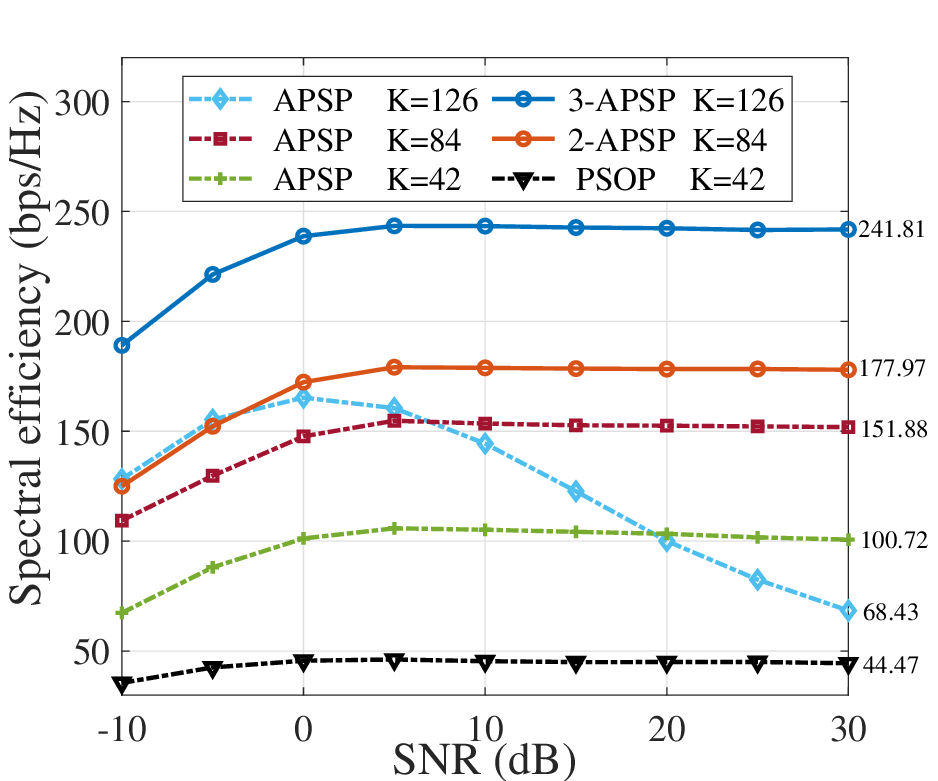}
			\captionsetup{font=footnotesize}
			\label{UMa_rate}}
		\subfigure[UMi]{
			\centering
			\includegraphics[width=0.31\linewidth]{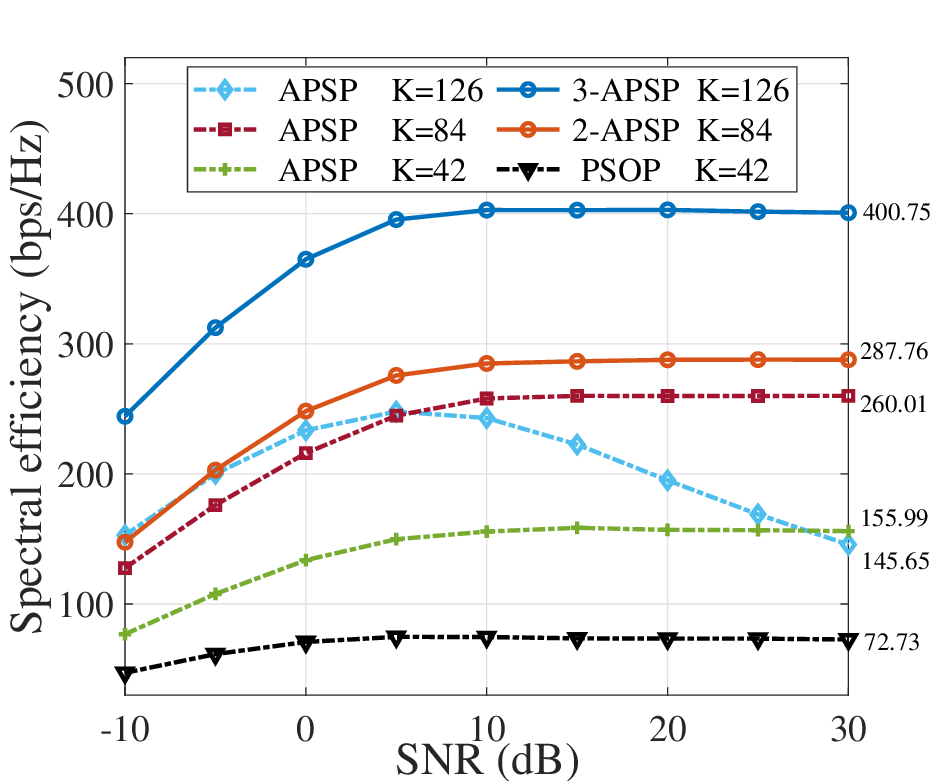}
			\captionsetup{font=footnotesize}
			\label{UMi_rate}}
		\subfigure[Indoor-office]{
			\centering
			\includegraphics[width=0.31\linewidth]{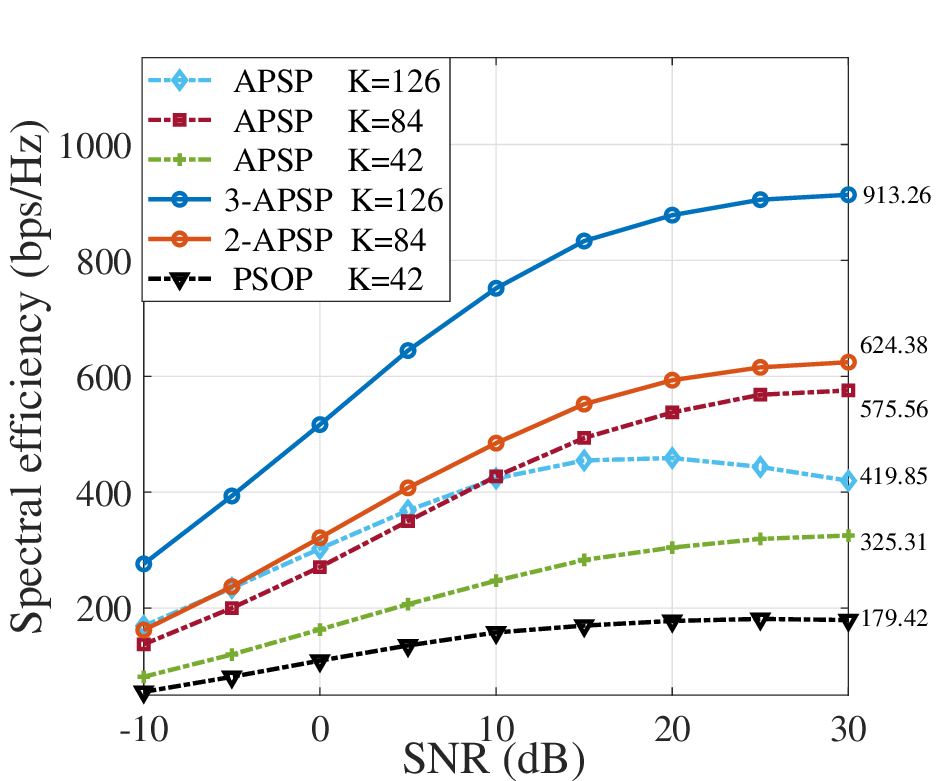}
			\captionsetup{font=footnotesize}
			\label{Id_rate}}
		
		\captionsetup{font=footnotesize}
		\caption{Comparison of the achievable spectral efficiency between the APSP and MAPSP methods under different UT numbers}
		\label{fig:rate}
	\end{figure*}
	
	According to the results illustrated in \figref{fig:MMSE}, the proposed MAPSP method demonstrates lower MMSE for the same number of UTs. This advantage becomes increasingly pronounced as $K$ increases. While the CS method achieves lower estimation error, its requirement of multiple OFDM symbols negatively affects SE in subsequent evaluations. In all scenarios, when $K=84$, the MMSE of 2-APSP is just slightly greater than the APSP at $K=42$. When $K=126$, the APSP method in \figref{UMa_MMSE} and \figref{UMi_MMSE} nearly fails in high-speed mobility conditions, while the MAPSP method maintains an acceptable estimation accuracy. In \figref{Id_MMSE}, the MMSE of 3-APSP is even slightly lower than the $K=84$ APSP. In summary, the MAPSP method exhibits significant MMSE performance as $K$ multiplicative increased, particularly in high mobility scenarios. 
	
	Next, we proceed to compare the SE of the proposed MAPSP in three different scenarios. The frame structure Type-B in \cite{you2015channel} is utilized, with each frame having a length of $500\mu$s, equivalent to 7 OFDM symbols \cite{3gpp36.211}. Each of the UL and DL data transmissions occupies half of the data segment, comprising 3 OFDM symbols each. A single OFDM symbol is devoted to pilot transmission, with the UL pilot segment positioned between the UL and DL data segments. This frame structure is well-suited for high mobility scenarios, and phase shift pilot methods encounter rapid channel changes felicitously. Both UL and DL data are transmitted using the MMSE receiver and precoder, assuming that the SNR is the same as that of the pilot SNR. By referring to the theoretical derivation in \cite{you2015channel}-\uppercase\expandafter{\romannumeral3}.C, we can infer the channel prediction by the MAPSP method as follows,
	\begin{equation}
		[\hat{\mathbf{H}}_{k'_{q'},\ell+\Delta_\ell}]_{i,j}=\varrho_{k'_{q'}}(\Delta_{\ell})\frac{[\mathbf{P}_{k'_{q'}}]_{i,j}[\mathbf{Y}_{k'_{q'},\ell}]_{i,j}}{\sum_{k_{q'}=0}^{K_{q'}-1}\left[\mathbf{P}_{k_{q'}}^{\varphi_{k_{q'}}-\varphi_{k'_{q'}}}\right]_{i,j}+\frac{1}{\eta _{\mathrm{tr}}}}.
	\end{equation}
	 And the MMSE error of channel prediction, 
	\begin{align}
		\sigma ^{\mathrm{pre}}_{k'_{q'}} \triangleq&\sum_{i=0}^{M-1}\sum_{j=0}^{N_{\mathrm{g}}-1}\mathsf{E}\left\{\left|\left[\mathbf{H}_{k'_{q'},\ell}\right]_{i,j}\right|^{2}-\left|\left[\hat{\mathbf{H}}_{k'_{q'},\ell+\Delta_\ell}\right]_{i,j}\right|^{2}\right\}. \nonumber\\
	\end{align}
	
	In pilot scheduling, the conditions for channel prediction and channel estimation are essentially identical. Specifically, the MMSE in channel prediction is minimized when pilot scheduling is performed based on the MMSE criterion of channel estimation. 
	
	According to \figref{fig:rate}, MAPSP exhibits relatively strong performance. Due to distinct UT channel sparsity patterns, CS cannot utilize sparsity as flexibly as APSP/MAPSP, requiring greater SE sacrifice for accurate estimation, resulting in inferior performance. The proposed MAPSP method significantly enhances SE compared to the APSP method when $K$ is the same. This improvement can be attributed to two factors affecting SE: the quantity of UT and the accuracy of channel estimation. $SE_{\mathrm{base}}$ denotes the SE when $K=42$ in each subfigure. For instance, in \figref{UMa_rate}, when $K=42$, SNR$=30$dB, the SE of 2-APSP experiences an increase of $177.97-100.72=0.77SE_{\mathrm{base}}$ compared to APSP. Building on this, 3-APSP shows an additional increase of $241.81-177.97=0.63SE_{\mathrm{base}}$. However, due to the greater inter-group pilot interference of 3-APSP, the improvement in SE is less pronounced. When $K=84$, the SE increase of APSP is $151.88-100.72=0.51SE_{\mathrm{base}}$ lower than that in 2-APSP due to an uptick in estimation errors stemming from increased intra-group channel overlaps. 
	
	Notably, at $K=126$, as the SNR rises, the impact of AWGN diminishes while the influence of pilot interference becomes more pronounced. At this stage, the APSP method is unable to accommodate all UTs, resulting in significant intra-group pilot interference that sharply reduces SE. This highlights the considerable potential of MAPSP in enhancing SE. In \figref{Id_rate}, the reduction in estimated performance due to the increase in UTs is less severe, resulting in a greater increment ratio for the MAPSP method, at $0.92SE_{\mathrm{base}}$ and $0.89SE_{\mathrm{base}}$, respectively. By comprehensively analyzing the UMa, UMi, and Indoor-office scenarios, when SNR = 30 dB and $K=84$, the SE gain of MAPSP over APSP is $17.2\%$, $10.7\%$, and $8.5\%$ respectively. When $K=126$, the gain becomes $253.4\%$, $175.2\%$, and $117.5\%$. This further demonstrates the superiority of MAPSP in the case of a large number of UTs and high mobility.
	
	\section{Conclusion}\label{sec_conclusion} 
	
	In this paper, we proposed the MAPSP method for acquiring a massive MIMO-OFDM channel. Based on the proposed system model, we investigated the channel estimation associated with MAPSP and conducted a detailed analysis of the principles underlying the generation and alleviation of pilot interference. Building on this foundation, we demonstrated the unity of two types of pilot interference within the same group and across different groups, and we established design criteria for optimal basic pilot matrix groups. Further, we proposed a MAPSP implementation that alleviates intra-group and inter-group interference by pilot scheduling and received signal pre-processing, respectively. The final simulation results showed that the proposed method significantly enhances estimation accuracy and SE in mobility scenarios. However, the MAPSP method's performance relies on prior statistical CSI availability and involves computationally intensive preprocessing. Future research may investigate efficient precoding schemes that exploit channel argument distribution properties, as well as potential applications in emerging scenarios such as vehicular networks and unmanned aerial vehicle formations.
	
	\appendices
	
	\section{Proof Of Formula (\ref{basic cro corr})}\label{appendix a}
	
	We choose the ZC sequence $\mathbf{s}_{q'}=\boldsymbol{\chi}_{N_\mathrm{c},0}^\mathfrak{r}$ of length $N_\mathrm{c}$, and the sequence $\mathbf{s}_{q}=\boldsymbol{\chi}_{N_\mathrm{c},\phi}^\mathfrak{r}$ after $\mathbf{s}_{q'}$ with cyclic shift $\phi$ as the basic pilot sequences. The corresponding basic pilot matrices are $\mathbf{S}_{q'}=\operatorname{diag}\left\{\mathbf{s}_{q'}\right\}$ and $\mathbf{S}_{q}=\operatorname{diag}\left\{\mathbf{s}_{q}\right\}$. Then, the SFPCM between $\mathbf{S}_{q}$ and $\mathbf{S}_{q'}$ is denoted as $\mathbf{R}^{q}_{q'}=\mathbf{S}_{q}\mathbf{S}_{q'}^{H}$. The $n\text{th}$ diagonal element of $\mathbf{R}^{q}_{q'}$ is denoted as
	\begin{equation}\label{ZC SFPCM}
		\begin{aligned}
			r_n=&\left[\boldsymbol{\chi}_{N_\mathrm{c},\phi}^\mathfrak{r}\right]_n\left[\boldsymbol{\chi}_{N_\mathrm{c},0}^{\mathfrak{r}\ *}\right]_n\\
			=&\exp\left(-\bar{\imath}\pi\frac{\mathfrak{r}\cdot \left(\mathscr{N}^2+\mathscr{N}-n^2-n\right)}{N_\mathrm{c}}\right),
		\end{aligned}
	\end{equation}
	where $\mathscr{N}$ is defined in (\ref{ZC mod}). 
	Substituting (\ref{ZC SFPCM}) into (\ref{basic ADPCM}), the ADPCM of the basic pilot matrix is
	\begin{align}\label{ZC corr ZC}
		\left[\mathbf{Z}^{q}_{q'}\right]_{i,j}=&\left[\mathbf{W}_{N_{\mathrm{c}}}^T\mathbf{R}^{q}_{q'}\mathbf{W}_{N_{\mathrm{c}}}^*\right]_{i,j}\nonumber\\
		=&\sum_{n=0}^{N_{\mathrm{c}-1}}r_n\cdot \exp\left(-\frac{\bar{\imath}2\pi(i-j)n}{N_{\mathrm{c}}}\right)\nonumber\\
		=&\sum_{n=0}^{N_{\mathrm{c}-1}}\left[\boldsymbol{\chi}_{N_\mathrm{c},\phi}^\mathfrak{r}\boldsymbol{\chi}_{N_\mathrm{c},0}^{\mathfrak{r}\ *}\right]_n\cdot \exp\left(-\frac{\bar{\imath}2\pi(i-j)n}{N_{\mathrm{c}}}\right)\nonumber\\
		=&\sum_{n=0}^{N_{\mathrm{c}-1}}\exp\left(-\bar{\imath}\pi\frac{\mathfrak{r}\cdot \left(\mathscr{N}^2+\mathscr{N}-n^2-n\right)}{N_\mathrm{c}}\right)\nonumber\\
		&\cdot \exp\left(-\frac{\bar{\imath}2\pi(i-j)n}{N_{\mathrm{c}}}\right).
	\end{align}
	
	For ease of analysis, we only consider the first column of $\mathbf{Z}^{q}_{q'}$, the complete ADPCM matrix can be derived further using the properties of the Toeplitz matrix. 
	
	Since $n-\phi \in [2-N_\mathrm{c},N_\mathrm{c}-1]$, the representation of $\mathscr{N}$ varies with the value of $n$,
	\begin{equation}
		\mathscr{N}=\begin{cases}
			n-\phi  &,\ n\ge\phi\\
			N_{\mathrm{c}}+n-\phi  &,\ n<\phi
		\end{cases}.
	\end{equation}
	Therefore, we divide (\ref{ZC corr ZC}) into two parts to sum, namely $n<\phi$ and $n\ge\phi$, defined as $\left[\mathbf{Z}^{q}_{q'}\right]_{i,j}=\varpi_{i,j}^{\mathrm{neg}}+\varpi_{i,j}^{\mathrm{pos}}$. The corresponding derivation is as follows.
	\begin{align}\label{neq sum}
		\varpi_{i,j}^{\mathrm{neg}}=&\sum_{n=0}^{\phi-1}\exp\left(\frac{-\bar{\imath}\pi}{N_\mathrm{c}}\mathfrak{r} \left(N_{\mathrm{c}}-\phi\right)\left(N_{\mathrm{c}}-\phi+2n+1\right)\right)\nonumber\\
		&\cdot \exp\left(-\frac{\bar{\imath}2\pi(i-j)n}{N_{\mathrm{c}}}\right)\nonumber\\
		=&\exp\left(\frac{-\bar{\imath}\pi}{N_\mathrm{c}}\mathfrak{r} \left(N_{\mathrm{c}}-\phi\right)\left(N_{\mathrm{c}}-\phi+1\right)\right)\cdot\Sigma^{\mathrm{neg}}.\\
		\label{pos sum}\varpi_{i,j}^{\mathrm{pos}}=&\sum_{n=\phi}^{N_{\mathrm{c}}-1}\exp\left(\frac{-\bar{\imath}\pi}{N_\mathrm{c}}\mathfrak{r} \phi\left(\phi-2n-1\right)\right)\nonumber\\
		&\cdot\exp\left(-\frac{\bar{\imath}2\pi(i-j)n}{N_{\mathrm{c}}}\right)\nonumber\\
		=&\exp\left(\frac{-\bar{\imath}\pi}{N_\mathrm{c}}\mathfrak{r}\phi\left(\phi-1\right)\right)\cdot\Sigma^{\mathrm{pos}},
	\end{align}
	where $\Sigma^{\mathrm{neg}}$ and $\Sigma^{\mathrm{pos}}$ represent the summations of the geometric sequence. For $\varpi_{i,j}^{\mathrm{neg}}$, the summation term $\Sigma^{\mathrm{neg}}$ in (\ref{neq sum}) is denoted as
	\begin{align}\label{Sigma neg}
		\Sigma^{\mathrm{neg}}=&\sum_{n=0}^{\phi-1}\exp\left(\frac{-\bar{\imath}2\pi}{N_\mathrm{c}}\left(\mathfrak{r}(N_\mathrm{c}-\phi)+(i-j)\right)n\right)\nonumber\\
		=&\exp\left(\frac{-\bar{\imath}\pi}{N_\mathrm{c}}\left(\mathfrak{r}(N_\mathrm{c}-\phi)+i-j\right)(\phi-1)\right)\nonumber\\
		&\cdot\frac{\sin \left(-\frac{\pi}{N_\mathrm{c}}\left(\mathfrak{r}(N_\mathrm{c}-\phi)+i-j\right)\phi\right)}{\sin \left(-\frac{\pi}{N_\mathrm{c}}\left(\mathfrak{r}(N_\mathrm{c}-\phi)+i-j\right)\right)}.
	\end{align}
	Intuitively, $\Sigma^{\mathrm{neg}}$ is not zero if and only if $\left\langle i-j\right\rangle_{N_\mathrm{c}}=\left\langle \mathfrak{r}\phi\right\rangle_{N_\mathrm{c}}$. Since this is the first column of $\mathbf{Z}^{q}_{q'}$, which means $0\le i-j\le N_\mathrm{c}-1$, so we have $i-j=\mathfrak{r}\phi$. Rewrite (\ref{Sigma neg}) 
	\begin{equation}
		\Sigma^{\mathrm{neg}}=\phi\cdot\exp\left(-\bar{\imath}\pi\mathfrak{r}(\phi-1)\right)\cdot\frac{\cos \left(-\pi\mathfrak{r}\phi\right)}{\cos \left(-\pi\mathfrak{r}\right)}=\phi.
	\end{equation}
	
	In a similar way, we can obtain $\Sigma^{\mathrm{pos}}$,
	\begin{align}\label{Sigma pos}
		\Sigma^{\mathrm{pos}}=&\sum_{n=\phi}^{N_{\mathrm{c}}-1}\exp\left(\frac{2\bar{\imath}\pi}{N_\mathrm{c}}\left(\mathfrak{r}\phi-i+j\right)n\right)\nonumber\\
		=&\exp\left(\frac{\bar{\imath}\pi}{N_\mathrm{c}}\left(\mathfrak{r}\phi+i-j\right)\left(N_\mathrm{c}+\phi-1\right)\right)\nonumber\\
		&\cdot\frac{\sin \left(\frac{\pi}{N_\mathrm{c}}\left(\mathfrak{r}\phi-i+j\right)\left(N_\mathrm{c}-\phi\right)\right)}{\sin \left(\frac{\pi}{N_\mathrm{c}}\left(\mathfrak{r}\phi-i+j\right)\right)}.
	\end{align}
	Reapplying the condition $i-j = \mathfrak{r}\phi$, we have
	\begin{equation}
	\Sigma^{\mathrm{pos}}=(N_\mathrm{c}-\phi)\cdot\frac{\cos \left(\frac{\pi}{N_\mathrm{c}}(\mathfrak{r}\phi-i+j)(N_\mathrm{c}-\phi)\right)}{\cos \left(\frac{\pi}{N_\mathrm{c}}(\mathfrak{r}\phi-i+j)\right)}=N_\mathrm{c}-\phi.
	\end{equation}
	
	Substituting (\ref{neq sum}) and (\ref{pos sum}) into (\ref{ZC corr ZC}), we have
	\begin{align}
		\left[\mathbf{Z}^{q}_{q'}\right]_{i,j}=&\exp\left(\frac{-\bar{\imath}\pi}{N_{\mathrm{c}}}\mathfrak{r}\phi(\phi-1)\right)\nonumber\\
		&\cdot\big(\phi\exp(-\bar{\imath}\pi\mathfrak{r}(N_{\mathrm{c}}-2\phi+1))+N_{\mathrm{c}}-\phi\big)\nonumber\\
		=&N_\mathrm{c}\exp\left(\frac{-\bar{\imath}\pi}{N_{\mathrm{c}}}\mathfrak{r}\phi(\phi-1)\right).
	\end{align}
	This completes the proof.
	
	\section{Proof Of Elimination Method}\label{appendix b}
	
	The role of the return-to-zero argument matrix is to adjust the mean of the argument distribution to zero. In other words, the mean arguments of $[\mathbf{\Theta}_{0}^{k_{q}}\!\odot\!\mathbf{H}_{k_{q},\ell}]_{i,j}$ are all zero. To simplify the proof, we assume that the arguments of all channels $\bar{\theta}\sim\mathcal{WN}(\bar{\mu},\bar{\sigma}^2)$ with $\bar{\mu}=0$. The mathematical expectation of the $n\text{th}$ column of channel vector for UT $k_q$ on the $\ell\text{th}$ OFDM symbol is
	\begin{align}\label{exp gaussian expect}
		\mathsf{E}\left\{\mathbf{h}_{k_q,\ell}^{n}\right\}
		=\sqrt{\left[\mathbf{P}_{k_q}\right]_{:,n}}\exp(\bar{\imath}\bar{\mu}-\frac{\bar{\sigma}^2}{2}).		
	\end{align}

	Since $\bar{\mu}=0$, we have 
	\begin{align}
		\mathsf{E}\left\{\mathbf{\Theta}_{0}^{k'_{q'}}\!\!\!\odot\!\!\left[\breve{\mathbf{Y}}_{k'_{q'},\ell}\right]_{i,j}\right\}=\mathsf{E}\left\{\left[\breve{\mathbf{Y}}_{k'_{q'},\ell}\right]_{i,j}\right\}.
	\end{align}
	The mathematical derivation as shown in (\ref{Expect_receiv_sig}) is obtained. This completes the proof.
	
		\begin{figure*}
			\begin{align}\label{Expect_receiv_sig}
				\mathsf{E}\left\{\left[\breve{\mathbf{Y}}_{k'_{q'},\ell}\right]_{i,j}\right\}=&\mathsf{E}\left\{\Re\left\{\frac{1}{N_{\mathrm{c}}}\left[\sum_{k_{q'}\in  \mathcal{K}_{q'}}\mathbf{H}_{k_{q'},\ell}^{\varphi_{k_{q'}}-\varphi_{k'_{q'}}}\right]_{i,j}\right\}\right.+\Re\left\{\frac{1}{N_{\mathrm{c}}}\left[\sum_{q\in\mathcal{Q}\backslash q'}\sum_{k_q=0}^{K_q-1}\mathbf{H}_{k_q,\ell}^{\varphi_{k_{q}}-\varphi_{k'_{q'}}}\right]_{i,j}\right\}+\frac{1}{\sqrt{\eta_{\mathrm{tr}} N_{\mathrm{c}}}}\left[\mathbf{N}_{\mathrm{nor}}\right]_{i,j}\nonumber\\
				&-\frac{1}{\tan\left[\mathbf{\Theta}_{\Sigma}^{q'}\right]_{i,j}}\left(\Im\left\{\frac{1}{N_{\mathrm{c}}}\left[\sum_{k_{q'}\in  \mathcal{K}_{q'}}\mathbf{H}_{k_{q'},\ell}^{\varphi_{k_{q'}}-\varphi_{k'_{q'}}}\right]_{i,j}\right\}\right.+\left.\left.\Im\left\{\frac{1}{N_{\mathrm{c}}}\left[\sum_{q\in\mathcal{Q}\backslash q'}\sum_{k_q=0}^{K_q-1}\mathbf{H}_{k_q,\ell}^{\varphi_{k_{q}}-\varphi_{k'_{q'}}}\right]_{i,j}\right\}\right)\right\}\nonumber\\			
				=&\mathsf{E}\left\{\cos\bar{\theta}\right\}\left[\sum_{k_{q'}\in  \mathcal{K}_{q'}}\sqrt{\mathbf{P}_{k_{q'},\ell}^{\varphi_{k_{q'}}-\varphi_{k'_{q'}}}}\right]_{i,j}+\mathsf{E}\left\{\cos(\bar{\theta}+\left[\mathbf{\Theta}_{\Sigma}^{q'}\right]_{i,j})\right\}\left[\sum_{q\in\mathcal{Q}\backslash q'}\sum_{k_q=0}^{K_q-1}\sqrt{\mathbf{P}_{k_q}^{\varphi_{k_{q}}-\varphi_{k'_{q'}}}}\right]_{i,j}\nonumber\\
				&+\mathsf{E}\left\{\frac{1}{\sqrt{\eta_{\mathrm{tr}} N_{\mathrm{c}}}}\left[\mathbf{N}_{\mathrm{nor}}\right]_{i,j}\right\}-\mathsf{E}\left\{\frac{\sin\bar{\theta}}{\tan\left[\mathbf{\Theta}_{\Sigma}^{q'}\right]_{i,j}}\right\}\left[\sum_{k_{q'}\in  \mathcal{K}_{q'}}\sqrt{\mathbf{P}_{k_{q'},\ell}^{\varphi_{k_{q'}}-\varphi_{k'_{q'}}}}\right]_{i,j}\nonumber\\
				&-\mathsf{E}\left\{\frac{\sin(\bar{\theta}+\left[\mathbf{\Theta}_{\Sigma}^{q'}\right]_{i,j})}{\tan\left[\mathbf{\Theta}_{\Sigma}^{q'}\right]_{i,j}}\right\}\left[\sum_{q\in\mathcal{Q}\backslash q'}\sum_{k_q=0}^{K_q-1}\sqrt{\mathbf{P}_{k_q}^{\varphi_{k_{q}}-\varphi_{k'_{q'}}}}\right]_{i,j}\nonumber\\
				=&\exp(-\frac{\bar{\sigma}^2}{2})\left[\sum_{k_{q'}\in  \mathcal{K}_{q'}}\sqrt{\mathbf{P}_{k_{q'},\ell}^{\varphi_{k_{q'}}-\varphi_{k'_{q'}}}}\right]_{i,j}+\exp(-\frac{\bar{\sigma}^2}{2})\cos\left[\mathbf{\Theta}_{\Sigma}^{q'}\right]_{i,j}\left[\sum_{q\in\mathcal{Q}\backslash q'}\sum_{k_q=0}^{K_q-1}\sqrt{\mathbf{P}_{k_q}^{\varphi_{k_{q}}-\varphi_{k'_{q'}}}}\right]_{i,j}\nonumber\\
				&-\frac{\exp(-\frac{\bar{\sigma}^2}{2})\sin\left[\mathbf{\Theta}_{\Sigma}^{q'}\right]_{i,j}}{\tan\left[\mathbf{\Theta}_{\Sigma}^{q'}\right]_{i,j}}\left[\sum_{q\in\mathcal{Q}\backslash q'}\sum_{k_q=0}^{K_q-1}\sqrt{\mathbf{P}_{k_q}^{\varphi_{k_{q}}-\varphi_{k'_{q'}}}}\right]_{i,j}=\exp(-\frac{\bar{\sigma}^2}{2})\left[\sum_{k_{q'}\in  \mathcal{K}_{q'}}\sqrt{\mathbf{P}_{k_{q'},\ell}^{\varphi_{k_{q'}}-\varphi_{k'_{q'}}}}\right]_{i,j}
			\end{align}
			\rule{18.2cm}{0.02cm}
		\end{figure*}
	
	\bibliographystyle{IEEEtran}
	\bibliography{EE_AI}

\begin{thebibliography}{10}
\providecommand{\url}[1]{#1}
\csname url@samestyle\endcsname
\providecommand{\newblock}{\relax}
\providecommand{\bibinfo}[2]{#2}
\providecommand{\BIBentrySTDinterwordspacing}{\spaceskip=0pt\relax}
\providecommand{\BIBentryALTinterwordstretchfactor}{4}
\providecommand{\BIBentryALTinterwordspacing}{\spaceskip=\fontdimen2\font plus
\BIBentryALTinterwordstretchfactor\fontdimen3\font minus
  \fontdimen4\font\relax}
\providecommand{\BIBforeignlanguage}[2]{{%
\expandafter\ifx\csname l@#1\endcsname\relax
\typeout{** WARNING: IEEEtran.bst: No hyphenation pattern has been}%
\typeout{** loaded for the language `#1'. Using the pattern for}%
\typeout{** the default language instead.}%
\else
\language=\csname l@#1\endcsname
\fi
#2}}
\providecommand{\BIBdecl}{\relax}
\BIBdecl

\bibitem{YZLY2025}
Y.~Zhao, L.~You, and et~al., ``Channel estimation in massive {MIMO-OFDM} with
  multi-group adjustable phase shift pilots,'' in \emph{Proc. IEEE 101st Veh.
  Technol. Conf. (VTC-Spring)}, Oslo, Norway, 2025, pp. 1--5.

\bibitem{HCLY2025}
H.~Che, L.~You, and et~al., ``Non-orthogonal pilot design exploiting spatial
  non-stationarity for {TDD} {XL-MIMO} transmission,'' \emph{IEEE Trans. Veh.
  Technol.}, vol.~74, no.~2, pp. 3424--3428, Feb. 2025.

\bibitem{ericsson2024report}
\BIBentryALTinterwordspacing
Ericsson, ``Mobile data traffic outlook,'' 2021. [Online]. Available:
  \url{https://www.ericsson.com/en/reports-and-papers/mobility-report/dataforecasts/mobile-traffic-forecast}
\BIBentrySTDinterwordspacing

\bibitem{ZJLY2025}
Z.~Jin, L.~You, and et~al., ``An {I2I} inpainting approach for efficient
  channel knowledge map construction,'' \emph{IEEE Trans. Wireless Commun.},
  vol.~24, no.~2, pp. 1415--1429, Feb. 2025.

\bibitem{TLM2015}
T.~L. Marzetta, ``Massive {MIMO}: An introduction,'' \emph{Bell Labs Tech. J.},
  vol.~20, pp. 11--22, Mar. 2015.

\bibitem{LYXC2020}
L.~You, X.~Chen, and et~al., ``Network massive {MIMO} transmission over
  millimeter-wave and terahertz bands: Mobility enhancement and blockage
  mitigation,'' \emph{IEEE J. Sel. Areas Commun.}, vol.~38, no.~12, pp.
  2946--2960, Dec. 2020.

\bibitem{YZhuLY2025}
Y.~Zhu, L.~You, and et~al., ``Robust precoding for massive {MIMO} {LEO}
  satellite integrated communication and localization systems,'' \emph{IEEE
  Commun. Lett.}, vol.~29, no.~1, pp. 21--25, Jan. 2025.

\bibitem{WCXL2023}
W.~Chen, X.~Lin, and et~al., ``5{G}-advanced toward 6{G}: Past, present, and
  future,'' \emph{IEEE J. Sel. Areas Commun.}, vol.~41, no.~6, pp. 1592--1619,
  Jun. 2023.

\bibitem{prasad2004ofdm}
R.~Prasad, \emph{{OFDM} for Wireless Communications Systems}, London, U.K.:
  Artech House, 2004.

\bibitem{SHLL2024}
S.~Hu, L.~Lian, and et~al., ``Blind multi-level {MAP} detection with phase
  noise compensation in {MIMO-OFDM} systems,'' \emph{IEEE Trans. Commun.},
  vol.~72, no.~3, pp. 1596--1611, Mar. 2024.

\bibitem{JZJZ2022}
J.~Zheng, J.~Zhang, and et~al., ``Cell-free massive {MIMO-OFDM} for high-speed
  train communications,'' \emph{IEEE J. Sel. Areas Commun.}, vol.~40, no.~10,
  pp. 2823--2839, Oct. 2022.

\bibitem{TMRT2022}
T.~Murakami, R.~Taniguchi, and et~al., ``Performance evaluation of uplink
  multiuser {MIMO-OFDM} system with single {RF} chain receiver,'' \emph{IEEE
  Access}, vol.~10, pp. 118\,878--118\,887, Oct. 2022.

\bibitem{XNLY2024}
X.~Niu, L.~You, and et~al., ``Coordinated multicast and unicast robust
  transmission in {C-V2V} massive {MIMO} with imperfect {CSI},'' \emph{IEEE
  Trans. Veh. Technol.}, vol.~73, no.~11, pp. 17\,106--17\,121, Nov. 2024.

\bibitem{LYXQ2024}
L.~You, X.~Qiang, and et~al., ``Integrated communications and localization for
  massive {MIMO} {LEO} satellite systems,'' \emph{IEEE Trans. Wireless
  Commun.}, vol.~23, no.~9, pp. 11\,061--11\,075, Sep. 2024.

\bibitem{dahlman20134g}
E.~Dahlman, S.~Parkvall, and J.~Skold, \emph{4G: LTE/LTE-advanced for mobile
  broadband}.\hskip 1em plus 0.5em minus 0.4em\relax Academic press, 2013.

\bibitem{XLWW2023}
X.~Liu, W.~Wang, and et~al., ``Structured hybrid message passing based channel
  estimation for massive {MIMO}-{OFDM} systems,'' \emph{IEEE Trans. Veh.
  Technol.}, vol.~72, no.~6, pp. 7491--7507, Jun. 2023.

\bibitem{TLNN2022}
T.~Li, N.~Noels, and et~al., ``Adaptive pilot allocation for estimating sparse
  uplink {MU}-{MIMO}-{OFDM} channels,'' \emph{IEEE Trans. Wireless Commun.},
  vol.~21, no.~10, pp. 8230--8244, Oct, 2022.

\bibitem{YZGS2022}
Y.~Zhu, G.~Sun, and et~al., ``{OFDM}-based massive grant-free transmission over
  frequency-selective fading channels,'' \emph{IEEE Trans. Commun.}, vol.~70,
  no.~7, pp. 4543--4558, Jul. 2022.

\bibitem{YLAS2025}
Y.~Li and A.~S. Madhukumar, ``Hybrid near- and far-field {TH}z {UM-MIMO}
  channel estimation: A sparsifying matrix learning-aided bayesian approach,''
  \emph{IEEE Trans. Wireless Commun.}, vol.~24, no.~3, pp. 1881--1897, Mar.
  2025.

\bibitem{LRMJ2024}
L.~Ribeiro and M.~Juntti, ``Spectral efficiency maximization for massive {MIMO}
  uplink with intra-cell pilot reuse,'' \emph{IEEE Wireless Commun. Lett.},
  vol.~13, no.~3, pp. 637--641, Mar. 2024.

\bibitem{LYXG2015}
L.~You, X.~Q. Gao, and et~al., ``Pilot reuse for massive {MIMO} transmission
  over spatially correlated rayleigh fading channels,'' \emph{IEEE Trans.
  Wireless Commun.}, vol.~14, no.~6, pp. 3352--3366, Jun. 2015.

\bibitem{YCLY2020}
Y.~Chen, L.~You, and et~al., ``Channel estimation with pilot reuse in {IQ}
  imbalanced massive {MIMO},'' \emph{IEEE Access}, vol.~8, pp. 1542--1555, Jan.
  2020.

\bibitem{LYMX2020}
L.~You, M.~Xiao, and et~al., ``Pilot reuse for vehicle-to-vehicle underlay
  massive {MIMO} transmission,'' \emph{IEEE Trans. Veh. Technol.}, vol.~69,
  no.~5, pp. 5693--5697, May 2020.

\bibitem{SSYC2020}
S.~Stein~Ioushua and Y.~C. Eldar, ``Pilot sequence design for mitigating pilot
  contamination with reduced {RF} chains,'' \emph{IEEE Trans. Commun.},
  vol.~68, no.~6, pp. 3536--3549, Jun. 2020.

\bibitem{XXBJ2016}
X.~Xiong, B.~Jiang, and et~al., ``{Q}o{S}-guaranteed user scheduling and pilot
  assignment for large-scale {MIMO}-{OFDM} systems,'' \emph{IEEE Trans. Veh.
  Technol.}, vol.~65, no.~8, pp. 6275--6289, Aug. 2016.

\bibitem{YCLY2021}
Y.~Chen, L.~You, and et~al., ``Channel estimation and robust detection for {IQ}
  imbalanced uplink massive {MIMO}-{OFDM} with adjustable phase shift pilots,''
  \emph{IEEE Access}, vol.~9, pp. 35\,864--35\,878, Mar. 2021.

\bibitem{you2015channel}
L.~You, X.~Q. Gao, and et~al., ``Channel acquisition for massive {MIMO}-{OFDM}
  with adjustable phase shift pilots,'' \emph{IEEE Trans. Signal Process.},
  vol.~64, no.~6, pp. 1461--1476, Mar. 2016.

\bibitem{JTXG2024}
J.~Tang, X.~Q. Gao, and et~al., ``Massive {MIMO}-{OFDM} channel acquisition
  with time-frequency phase-shifted pilots,'' \emph{IEEE Trans. Commun.},
  vol.~73, no.~6, pp. 4520--4535, Jun. 2024.

\bibitem{JWSL2019}
J.~W. Browning, S.~L. Cotton, and et~al., ``The rician complex envelope under
  line of sight shadowing,'' \emph{IEEE Commun. Lett.}, vol.~23, no.~12, pp.
  2182--2186, Dec. 2019.

\bibitem{JWSL2023}
------, ``A unification of {L}o{S}, {N}on-{L}o{S}, and {Q}uasi-{L}o{S} signal
  propagation in wireless channels,'' \emph{IEEE Trans. Antennas Propag.},
  vol.~71, no.~3, pp. 2682--2696, Mar. 2023.

\bibitem{SL2014}
S.~L. Cotton, ``A statistical model for shadowed body-centric communications
  channels: Theory and validation,'' \emph{IEEE Trans. Antennas Propag.},
  vol.~62, no.~3, pp. 1416--1424, Mar. 2014.

\bibitem{DWJF2011}
D.~W. Matolak and J.~Frolik, ``Worse-than-{R}ayleigh fading: Experimental
  results and theoretical models,'' \emph{IEEE Commun. Mag.}, vol.~49, no.~4,
  pp. 140--146, Apr. 2011.

\bibitem{WKIG2016}
W.~Khawaja, I.~Guvenc, and D.~Matolak, ``{UWB} channel sounding and modeling
  for {UAV} air-to-ground propagation channels,'' in \emph{Proc. - IEEE Glob.
  Commun. Conf., GLOBECOM}, Washington, DC, United states, 2016, pp. 1--7.

\bibitem{XNLY2020}
X.~Niu, L.~You, and X.~Gao, ``Phase shift adjustable pilots for channel
  acquisition in {V}ehicle-to-{V}ehicle underlay wideband massive {MIMO},''
  \emph{IEEE Access}, vol.~8, pp. 203\,793--203\,803, Nov. 2020.

\bibitem{MGNZ2023}
M.~Ghermezcheshmeh and N.~Zlatanov, ``Parametric channel estimation for {L}o{S}
  dominated holographic massive {MIMO} systems,'' \emph{IEEE Access}, vol.~11,
  pp. 44\,711--44\,724, May 2023.

\bibitem{THDC2018}
T.~Haelsig, D.~Cvetkovski, and et~al., ``Statistical properties and variations
  of {LOS} {MIMO} channels at millimeter wave frequencies,'' in \emph{WSA -
  Int. ITG Workshop Smart Antennas}, Bochum, Germany, 2018.

\bibitem{MP2011}
M.~Ptzold, \emph{Mobile radio channels}.\hskip 1em plus 0.5em minus 0.4em\relax
  2nd ed. Chichester, U.K.: Wiley, 2012.

\bibitem{WJDC1994}
W.~C. Jakes and D.~C. Cox, \emph{Microwave mobile communications}.\hskip 1em
  plus 0.5em minus 0.4em\relax Ed. New York, NY, USA: IEEE, 1994.

\bibitem{YCLY2021J}
Y.~Chen, L.~You, and et~al., ``Widely-linear processing for the uplink of the
  massive {MIMO} with {IQ} imbalance: Channel estimation and data detection,''
  \emph{IEEE Trans. Signal Process.}, vol.~69, pp. 4685--4698, Aug. 2021.

\bibitem{SDSLF2023}
D.~Shi, L.~Song, and et~al., ``Channel acquisition for {HF} skywave massive
  {MIMO}-{OFDM} communications,'' \emph{IEEE Trans. Wireless Commun.}, vol.~22,
  no.~6, pp. 4074--4089, Jun. 2023.

\bibitem{CSXG2015}
C.~Sun, X.~Gao, and et~al., ``Beam division multiple access transmission for
  massive {MIMO} communications,'' \emph{IEEE Trans. Commun.}, vol.~63, no.~6,
  pp. 2170--2184, Jun. 2015.

\bibitem{CWSJ2015}
C.-K. Wen, S.~Jin, and et~al., ``Channel estimation for massive {MIMO} using
  gaussian-mixture bayesian learning,'' \emph{IEEE Trans. Wireless Commun.},
  vol.~14, no.~3, pp. 1356--1368, Mar. 2015.

\bibitem{JTLY2025}
J.~Tang, L.~You, and et~al., ``Statistical {CSI} acquisition for
  multi-frequency massive {MIMO} systems,'' \emph{IEEE Trans. Commun.}, early
  access. 2025.

\bibitem{3gpp36.211}
3GPP, ``{3rd Generation Partnership Project; Technical Specification Group
  Radio Access Network; Evolved Universal Terrestrial Radio Access(E-UTRA);
  Physical Channels and Modulation (Release 12)},'' {3rd Generation Partnership
  Project (3GPP)}, Technical Specification (TS) 36.211, Mar. 2017, v12.9.0.

\bibitem{quadriga}
S.~Jaeckel, L.~Raschkowski, and et~al., ``{Q}ua{DR}i{G}a: A 3-{D} multi-cell
  channel model with time evolution for enabling virtual field trials,''
  \emph{IEEE Trans. Antennas Propag.}, vol.~62, no.~6, pp. 3242--3256, Jun.
  2014.

\bibitem{3gpp38.901}
3GPP, ``{3rd Generation Partnership Project; Technical Specification Group
  Radio Access Network; Study on channel model for frequencies from 0.5 to 100
  GHz (Release 14)},'' {3rd Generation Partnership Project (3GPP)}, Technical
  Report (TR) 38.901, Dec. 2017, v14.3.0.

\end{thebibliography}

\end{document}